\documentclass{birkjour}
 \newtheorem{thm}{Theorem}[section]

 \theoremstyle{definition}
 
 \theoremstyle{remark}

 \numberwithin{equation}{section}

\usepackage{graphicx}

\newcommand{\rmd}{\mathrm{d}}
\newcommand{\rme}{\mathrm{e}}

\begin{document}

%
%
%
%
%
%
%
%
%

\title[Examples of Naked Singularity Formation in Einstein Vacua]{Examples of Naked Singularity Formation in Higher-Dimensional Einstein-Vacuum Spacetimes}

\author[Xinliang An]{Xinliang An}

\address{
Department of Mathematics \\
University of Toronto \\
Toronto, M5S 2E4, Canada \\
Department of Mathematics \\ 
Rutgers University \\ 
Piscataway, NJ 08854-8019, USA
}
\email{xinliang.an@utoronto.ca}

\author{Xuefeng Zhang}

\address{
TianQin Reseach Center for Gravitational Physics \br
School of Physics and Astronomy \br
Sun Yat-sen University \br
Zhuhai, 519082, China \br
Departments of Physics and Astronomy \br 
Beijing Normal University \br
Beijing 100875, China
}
\email{zhangxf38@sysu.edu.cn}


\date{November 6, 2017}

\begin{abstract}
The vacuum Einstein equations in 5+1 dimensions are shown to admit solutions describing naked singularity formation in gravitational collapse from nonsingular asymptotically locally flat initial data that contain no trapped surface. We present a class of specific examples with topology $\mathbb{R}^{3+1} \times S^2$. Thanks to the Kaluza-Klein dimensional reduction, these examples are constructed by lifting continuously self-similar solutions of the 4-dimensional Einstein-scalar field system with a negative exponential potential. The latter solutions are obtained by solving a 3-dimensional autonomous system of first-order ordinary differential equations with a combined analytic and numerical approach. Their existence provides a new test-bed for weak cosmic censorship in higher-dimensional gravity. In addition, we point out that a similar attempt of lifting Christodoulou's naked singularity solutions of massless scalar fields fails to capture formation of naked singularities in 4+1 dimensions, due to a diverging Kretschmann scalar in the initial data.
\end{abstract}

\maketitle
\section{Introduction}

In the process of gravitational collapse, singularities can emerge under a broad range of circumstances. They are either covered by black hole horizons, or being visible to far-away observers---commonly referred to as naked singularities. In regard to these two outcomes, the weak cosmic censorship conjecture \cite{Penrose69} asserts that generically all singularities should be hidden inside black holes. During the past few decades, only limited progress has been made toward a general proof or disproof of the conjecture (see recent reviews \cite{Christo99,Dafermos09,Wald97,Penrose99,Rendall05}). In a series of papers from 1980s to late 90s, Christodoulou investigated the global dynamics of the Einstein-scalar field equations with a vanishing potential under the assumption of spherical symmetry. Remarkably, he established the first example of naked singularities that can evolve from regular asymptotically flat initial data in a matter field deemed physically suitable \cite{Christo94}. Later on, he also showed that such occurrences are unstable and non-generic \cite{Christo99}, and proved a version of weak cosmic censorship for the model system. Despite huge effort, the validity of the conjecture in general situations still remains elusive, which constitutes possibly the most important open problem in classical general relativity. Nevertheless, on a different path, significant progress has been made in mathematical studies of the formation of trapped surfaces in vacuum spacetimes without symmetry assumptions, notably in the recent monumental work by Christodoulou \cite{Christo09} and related development \cite{An12,An17,Dafermos13,Klainerman12,Klainerman14,Li14,Luk13,Reiterer11}.

By construction, Christodoulou's naked singularity solution in \cite{Christo94} is continuously self-similar and has \emph{two} real parameters $k$ ($\kappa$ in our notation) and $a$. Detailed analysis was given to the moment when the curvature singularity first appears along with the self-similar coordinate becoming singular as well. A naked singularity can form provided $0<k^2<1/3$. The asymptotically flat initial data sets are obtained by truncating the self-similar ones. Relevant to this work, Brady examined the same self-similar model using numerical calculation \cite{Brady95} and discussed its critical behavior (for an even earlier numerical attempt before \cite{Christo94}, see \cite{Goldwirth87}). Bedjaoui et al. further applied the same self-similar construction to the Brans-Dicke theory \cite{Bedjaoui10} and showed that naked singularities can also emerge. More examples of continuously self-similar solutions can be found, for instance, in \cite{Bizon00,Hirschmann04,Rendall11,Wang03}.

Further understanding of naked singularity formation in the Einstein-massless-scalar field was provided by Choptuik \cite{Choptuik93}. Though numerical simulations, he investigated various one-parameter families of initial data that result in either dispersal or collapse to a black hole, depending on the strength of incoming scalar-wave packets. Near the threshold of black hole formation, the black hole mass obeys a power-law scaling, and the spacetime always asymptotically approaches a universal solution that possesses discrete self-similarity. The finding indicates that the threshold solution corresponds to naked singularity formation \cite{Gundlach97,Martin03} that has a nature similar to Christodoulou's solution \cite{Wald97}. Since his work, similar behaviors, now called critical phenomena, have been reported in many other systems (see the recent review \cite{Gundlach07}), for instance, the Einstein-scalar field system with nontrivial self-interaction potentials (quadratic or double-well forms \cite{Brady97,Hawley00,Honda02}), and axisymmetric collapse of gravitational waves \cite{Abrahams93}. For more discussion of self-similarity in general relativity, see, e.g., \cite{Carr99,Carr05} and references therein. 

In parallel to advancement in 4 dimensions, string theory and brane-world models \cite{Maartens10} have greatly promoted the study of general relativity in higher dimensions. By entering dimensions higher than 4, novel topologies and horizon geometries can arise, easily evading Birkhoff's theorem and permitting a large variety of black objects \cite{Emparan08,Horowitz12}. Particularly on the dynamical aspect of the vacuum Einstein equations, critical phenomenon \cite{Choptuik93,Gundlach07} was reported in the collapse of 5-dimensional gravitational waves (Bianchi IX) through numerical calculation \cite{Bizon05}. At the threshold of black hole formation, the critical evolution is expected (yet to be demonstrated) to develop a naked singularity. Again in 5 dimensions, numerical simulations of thin black strings and black rings that are subject to the Gregory-Laflamme instability \cite{Gregory93} have also indicated formation of naked singularities in the final stage of a self-similar cascade \cite{Figueras16,Lehner10}. Moreover in 6 dimensions and by numerical methods, axisymmetric ultraspinning Myers-Perry black holes have been shown to evolve into naked singularities \cite{Figueras17}. Clearly, higher-dimensional spacetimes allow more possibilities that weak cosmic censorship can be challenged. 

In higher-dimensional theories, a powerful way to construct solutions is by dimensional reduction \cite{Duff86,Overduin97}. Originally proposed by Kaluza and Klein, the procedure assumes the spacetime to be the product of a 4-dimensional Lorentzian manifold and a homogeneous space representing extra dimensions. It further requires that all coordinate dependence of extra dimensions be dropped. With this approach, dimensional reduction of the vacuum Einstein equations leads to 4-dimensional Einstein gravity coupled to a scalar field with or without an exponential potential, depending on the geometry of the extra dimensions. In a similar fashion, exponential potentials (Liouville type) often arise from scale-invariant supergravity theories \cite{Townsend01,Duff86,Overduin97,Halliwell87}. They have also been intensively studied in scalar-field cosmology \cite{Halliwell87,Coley03}. Referring back to the dimensional reduction of vacuum, in the simpler case of a vanishing potential, i.e., the massless scalar field in 4 dimensions that has been investigated by Christodoulou, it is clear that one can further explore his results on naked singularity formation \cite{Christo94} from the perspective of higher dimensions.

In the context of anti-de Sitter/conformal field theory correspondence (AdS/CFT), there has been a series of discussions on possible violation of cosmic censorship in asymptotically AdS spacetimes \cite{Bizon11,Hertog04,Hertog04a,Garfinkle04,Gutperle04,Frolov04,Dafermos05,Langfelder05}. Most attention has focused on the Einstein-scalar field system with a scalar potential. Unfortunately, for non-vanishing potentials (possibly negative, bounded or unbounded from below), the question of weak cosmic censorship in spherical symmetry remains open \cite{Dafermos05}. Examples of naked singularity formation bearing similar strength to Christodoulou's appear to be lacking.

In this paper, we will first extend Christodoulou's self-similar model of naked singularities to include a positive or negative scalar potential of exponential forms, i.e., $V(\phi)=\pm\exp(2\phi/\kappa)$ with a parameter $\kappa$. The choice of exponential forms is crucial because of their compatibility with continuous self-similarity, which makes our attempt viable in the first place. In presenting our model, we will follow Brady's precedent of \cite{Brady95} with certain modifications (local compactification of the phase space) in order to exhibit the naked singularity solutions more prominently in numerical plots. When it comes to analysis of the equations and spacetimes, Christodoulou's paradigm is respected in our treatment. In particular, we can embed his 2-dimensional solution manifold into our 3-dimensional manifold, allowing us to draw analogy along the way.

Next we will consider the uplift of these 4-dimensional naked singularity solutions and decide whether or not their higher-dimensional descendants can represent formation of naked singularity in vacuum gravitational collapse. In particular, we will point out certain complications (loss of regularity) when such inheritance can be broken. Our objective is to show the following main result: \\
\centerline{
\begin{minipage}[c][2.5cm]{0.95\textwidth}
  \emph{In 5+1 dimensional spacetimes with the spatial symmetry of $S^2\times S^2$ orthogonal to a radial direction, and governed by the vacuum Einstein equations $\hat{R}_{\mu\nu}=0$, there exist nonsingular asymptotically locally flat initial data sets, free of trapped surfaces, which lead to naked singularity formation. }
\end{minipage}}
Unlike the situation in 5-dimensional black strings and black rings \cite{Figueras16,Lehner10}, the time evolution we consider does not begin with a pre-existing black object. Our result can be viewed as a first step towards constructing further analytic examples for understanding the issue of naked singularity formation in Einstein-vacuum spacetimes.

The paper is structured as follows. In Sec. \ref{sec:KK}, we introduce the self-gravitating scalar model by performing the dimensional reduction with a higher-dimensional metric ansatz. In Sec. \ref{sec:EOM}, under spherical symmetry and continuous self-similarity, we reduce the Einstein equations to a 3-dimensional autonomous system of evolutionary equations. They automatically incorporate the equations for the massless scalar field as a special case. Brady's notation in \cite{Brady95} is used for an easier comparison. In Sec. \ref{sec:local}, local analytic studies are carried out at critical and singular points of the phase space. Then in Sec. \ref{sec:global}, we piece local pictures together and numerically solve the equations in two parameter regimes $\kappa>1$ and $0<\kappa<1$, respectively. Afterwards, Sec. \ref{sec:4D} is devoted to analyzing the spacetime structure of the naked singularity solutions in 4 dimensions. In Sec. \ref{sec:6D}, for the negative exponential potential, we examine the 6-dimensional lifted solutions from the aspects of apparent horizons, the Kretschmann scalar, asymptotic behavior, strength of singularity, and homothetic symmetry. Sec. \ref{sec:5D} comments on a similar procedure for Christodoulou's solutions promoted to 5 dimensions. Concluding remarks are made in Sec. \ref{sec:conclu}. For readers interested in Christodoulou's paper \cite{Christo94}, the Appendix \ref{app:Christo} contains our autonomous equations using his original notation. Regarding the conceptual framework involved in this work, one may refer to \cite{Dafermos08} (Chs. 1 \& 2) for a quick survey of general relativity from the point of view of the Cauchy problem.


\section{Kaluza-Klein dimensional reduction} \label{sec:KK}

The Lagrangian (hatted) of pure Einstein gravity in $4+n$ dimensions is
\begin{equation} \label{LagHD}
 \hat{\mathcal{L}} = \sqrt{-\hat{g}} \hat{R}.
\end{equation}
Now introduce a warped-product metric ansatz
\begin{equation} \label{metric4n}
 \rmd \hat{s}^2 = \exp\!\bigg(2\sqrt{\frac{n}{n+2}}\, \phi \bigg) \rmd s^2
 + \exp\!\bigg(\frac{-4\phi}{\sqrt{n(n+2)}} \bigg) \rmd s_n^2,
\end{equation}
where $\rmd s^2$ (unhatted) and $\rmd s_n^2$ denote, respectively, the metric of 4-dimensional spacetime and the $n$-dimensional compactifying space $\mathcal{K}$. Both metrics are re-scaled by exponential functions of a scalar $\phi$ (dilaton). We assume that the scalar function $\phi$ only depend on the coordinates of $\rmd s^2$ and that the space $\mathcal{K}$ has a constant Ricci scalar $R_n$. By virtue of consistent Kaluza-Klein dimensional reduction on $\mathcal{K}$ \cite{Mignemi89,Bremer99,Cai04,Duff86,Overduin97}, one can verify that the vacuum Einstein equations as derived from (\ref{LagHD}) correspond to the equations of motion of the following effective action on $\rmd s^2$ and $\phi$ (see the Appendix \ref{app:EVE6} for an explicit calculation used in Sec. \ref{sec:6D}):
\begin{equation} \label{Lag4D}
 \mathcal{L} = \sqrt{-g} \bigg(\frac{R}{4} - \frac{1}{2}(\partial\phi)^2 + \frac{R_n}{4} \exp\!\bigg(2\sqrt{\frac{n+2}{n}} \phi\bigg) \bigg).
\end{equation}
The reduced action represents 4-dimensional Einstein gravity minimally coupled to a real scalar field with an exponential potential. Hence from the viewpoint of solving the Einstein equations under (\ref{metric4n}), the effect of having extra dimensions is tantamount to actuating a scalar field in 4-dimensional Einstein gravity, of which the self-interaction potential depends on the geometry of $\mathcal{K}$.

\section{The field equations} \label{sec:EOM}

In solving the reduced system (\ref{Lag4D}), it benefits to consider a slight variant:
\begin{align}
 \mathcal{L} &= \sqrt{-g} \left(\frac{R}{4} - \frac{1}{2}g^{\mu\nu}\nabla_\mu \phi \nabla_\nu \phi - V(\phi)\right), \label{Lag} \\
 V(\phi) &= V_0 \exp\!\left(\frac{2\phi}{\kappa}\right), \qquad \kappa\neq 0,
 \qquad V_0=\pm 1\ \text{or } 0,
 \label{Vphi}
\end{align}
where the potential $V(\phi)$ is parameterized by a real number $\kappa$. Values of $\kappa$ will be specified when one lifts the metric $\rmd s^2$ to higher dimensions via (\ref{metric4n}). By a translation of $\phi$, one can always normalize the multiplicative factor $V_0\neq 0$ to $\pm 1$. The energy-momentum tensor is given by
\begin{equation} \label{T}
 T_{\mu\nu} = \nabla_\mu \phi \nabla_\nu \phi - \left[\frac{1}{2}\left(\nabla_\alpha \phi \nabla^\alpha \phi\right) + V(\phi)\right] g_{\mu\nu}.
\end{equation}
The Einstein equations ($G_{\mu\nu}=2T_{\mu\nu}$) and the generalized Klein-Gordon equation are
\begin{align}
 R_{\mu\nu} &= 2\big(\nabla_\mu \phi \nabla_\nu \phi + V(\phi)g_{\mu\nu}\big), \label{Eeq} \\
 \nabla_\alpha \nabla^\alpha \phi &= V'(\phi). \label{KGeq}
\end{align}
Using retarded Bondi coordinates in spherical symmetry, we consider the following metric:
\begin{equation} \label{metric4}
 \rmd s^2 = -g(u,r) \tilde{g}(u,r)\,\rmd u^2 - 2 g(u,r)\,\rmd u\rmd r + r^2 \rmd\Omega^2,
\end{equation}
with $r$ the area radius and $\rmd\Omega^2$ the standard metric of a unit 2-sphere.
Then the field equations and the wave equation for $\phi=\phi(u,r)$ can be written as
\begin{align}
 (\ln g)_{,r} &= r (\phi_{,r})^2, \label{feq1} \\
 (r \tilde{g})_{,r} &= \left(1 - 2r^2 V(\phi)\right) g, \label{feq2} \\
 g(\tilde{g}/g)_{,u} &= 2r \left[(\phi_{,u})^2 - \tilde{g} \phi_{,u}\phi_{,r}\right], \label{feq3} \\
 r^{-1} (r^2 \tilde{g}\phi_{,r})_{,r} &= 2\phi_{,u} + 2r\phi_{,ru} + r g V'(\phi). \label{feq4}
\end{align}
For the center $r=0$ to be regular (see, e.g., \cite{Christo94}), we assume
\begin{equation} \label{BV}
 g(u,0) = \tilde{g}(u,0) = 1,
\end{equation}
which also fixes the null coordinate $u$ as the proper time of an observer at the center. Additionally, the Misner-Sharp (Hawking) mass is defined by
\begin{equation} \label{mass}
 1-\frac{2m(u,r)}{r} = g^{\mu\nu} r_{,\mu} r_{,\nu} = \frac{\tilde{g}}{g}.
\end{equation}

To construct self-similar solutions, we notice that the system (\ref{feq1}-\ref{feq4}) is invariant under the following scaling transformation:
\begin{align}
 & g \rightarrow g, \qquad \tilde{g} \rightarrow \tilde{g}, \qquad
 r \rightarrow a r, \qquad u \rightarrow a u, \qquad a>0, \nonumber \\
 & \phi \rightarrow \phi-\kappa \ln a, \qquad V(\phi) \rightarrow V(\phi)/a^2,
 \qquad \rmd s^2 \rightarrow a^2\rmd s^2.
\end{align}
It allows us to adopt the same self-similar ansatz as derived for the massless scalar field (see \cite{Christo94} and the Appendix A of \cite{Brady95} for more detailed derivation; one may jump ahead to Fig. \ref{fig4} for a graphical representation of the self-similar coordinate system):
\begin{align}
 \rmd s^2 &= -g(x) \tilde{g}(x)\, \rmd u^2 - 2 g(x)\, \rmd u\rmd r + r^2 \rmd\Omega^2, \label{metricx} \\
 \phi &= \bar{h}(x) - \kappa \ln(-u), \qquad x=-\frac{r}{u}, \qquad u<0,
\end{align}
with $\bar{h}$ a real-valued unknown to be determined. Here the metric (\ref{metricx}) admits a homothetic Killing vector ($L_\xi g_{\mu\nu} = 2g_{\mu\nu}$)
\begin{equation}
 \xi = u \partial_u + r \partial_r
\end{equation}
such that $\xi g(x) = 0 = \xi \tilde{g}(x)$ and $\xi \phi = -\kappa$. Apparently, the self-similar coordinate $x$ becomes singular at $u=0$, and in this paper we are primarily concerned with the past region $u<0$ as $u$ increases toward the future. Moreover, note that $\kappa$ also appears in the ansatz, and hence is fixed by the Lagrangian if $V_0\neq 0$. This arrangement is necessary in order to reduce the field equations to ordinary differential equations (ODEs).

Derived from (\ref{feq1}-\ref{feq4}), the equations for the four unknowns---$g(x)$, $\tilde{g}(x)$, $\bar{h}(x)$, and $\gamma(x)$ ($\gamma$ defined in (\ref{fode1}))---consist of four first-order ODEs ($\ '=\rmd/\rmd x$) plus one algebraic equation:
\begin{align}
 x \bar{h}' &= \gamma, \label{fode1} \\
 x g' &= g\gamma^2, \\
 x\tilde{g}' + \tilde{g} &= \big(1 - 2V_0 x^2 \rme^{2\bar{h}/\kappa}\big)\, g, \\
 (\tilde{g} - 2x)\, (x\gamma') &= 2\kappa x - (g - 2x)\, \gamma + 2V_0 x^2 \rme^{2\bar{h}/\kappa} g (\gamma + \kappa^{-1}), \label{fode4} \\
 g - \tilde{g} &= 2\kappa^2 x - (\tilde{g} - 2x)(\gamma^2 + 2\kappa\gamma) + 2V_0 x^2 \rme^{2\bar{h}/\kappa} g.
\end{align}
One may check that the derivative of the last equation which derives from (\ref{feq3}), is a combination of (\ref{fode1}-\ref{fode4}). If $V_0=0$, the equations simplify to the special case for massless scalar fields (see \cite{Brady95}, (2.9-2.13)).

To convert the above equations into an autonomous system, we follow Brady's method \cite{Brady95} and introduce new unknown variables:
\begin{equation} \label{wyzs}
 w(s) = x^2 \rme^{2\bar{h}/\kappa} \geq 0, \qquad y(s) = \frac{\tilde{g}}{g},
 \qquad z(s) = \frac{x}{\tilde{g}}, \qquad x = \rme^s \geq 0.
\end{equation}
Thus we obtain ($\ \dot{} = \rmd/\rmd s = x\rmd/\rmd x$; cf. \cite{Brady95}, (2.18-2.21))
\begin{align}
 \dot{w} &= 2w (\gamma + \kappa)/\kappa, \label{dew} \\
 \dot{y} &= 1 - y\,(1+\gamma^2) - 2V_0 w, \label{dey} \\
 \dot{z} &= z\left[2 - \left(1-2V_0 w \right)y^{-1}\right], \label{dez} \\
 (1-2z)\dot{\gamma} &= 2\kappa z - \gamma(y^{-1}-2z) + 2V_0 w y^{-1} (\gamma+\kappa^{-1}), \label{degamma} \\
 (1-2z)(\gamma+\kappa)^2 &= 1 + \kappa^2 - (1 - 2V_0 w) y^{-1}, \qquad V_0=\pm 1\ \text{or } 0.
 \label{eqa}
\end{align}
It can be verified that (\ref{dey}-\ref{eqa}) imply (\ref{dew}). By using the algebraic equation (\ref{eqa}), we can remove the term $V_0 w$ in (\ref{dey}-\ref{degamma}) and reduce the problem to a 3-dimensional dynamical system:
\begin{align}
 \dot{y} &= 2(\gamma+\kappa)\left[(\gamma+\kappa)z-\gamma\right] y, \label{dey3} \\
 \dot{z} &= -2(\gamma+\kappa)^2 z^2 + \left[(\gamma+\kappa)^2+1-\kappa^2\right] z, \label{dez3} \\
 (1-2z)\dot{\gamma} &= (1-2z)\left[\gamma^3 + \left(\frac{1}{\kappa}+2\kappa\right)\gamma^2\right]
 + (1-2z-2\kappa^2 z)\gamma - \frac{1}{\kappa}\left(1 - \frac{1}{y}\right). \label{degamma3}
\end{align}
Given these equations, $w$ as determined by (\ref{eqa}) satisfies (\ref{dew}). Other ways to bring forth autonomous systems are also possible. For instance, see the Appendix \ref{app:Christo} for comparison with Christodoulou's treatment and notation.

The system (\ref{dey3}-\ref{degamma3}), derived from (\ref{dew}-\ref{eqa}), treats solutions for $V_0=\pm 1$ on an equal footing, and also automatically includes Brady's special case for the massless scalar field, i.e., (\ref{dew}-\ref{eqa}) with $w=0$ (see also \cite{Brady95}, (2.18-2.20)). One can retrieve this latter case from (\ref{dey3}-\ref{degamma3}) by imposing the algebraic constraint (\ref{eqa}) with $V_0 w=0$.

To facilitate analysis of the autonomous system, we further introduce
\begin{equation} \label{zzeta}
 z = \frac{\zeta}{2(1-\zeta)}.
\end{equation}
This substitution is intended to bring $z=+\infty$ to a finite region (compactification), i.e., $\zeta=1$, as we will see later that naked singularity solutions tend to $z=+\infty$ as $s\rightarrow +\infty$. In terms of $y$, $\zeta$, and $\gamma$, the system (\ref{dey3}-\ref{degamma3}) takes the form
\begin{align}
 \dot{y} &= \frac{y(\gamma+\kappa)(3\gamma\zeta - 2\gamma + \kappa\zeta)}{1-\zeta}, \label{ode1} \\
 \dot{\zeta} &= -\left[2(\gamma+\kappa)^2 + 1 - \kappa^2\right] \zeta^2 + \left[(\gamma+\kappa)^2+1-\kappa^2\right] \zeta, \label{ode2} \\
 \dot{\gamma} &= \gamma^3 + \left(\kappa^{-1} + 2\kappa\right)\gamma^2
 + \left(1 - \frac{\kappa^2 \zeta}{1-2\zeta}\right)\!\gamma - \frac{1-\zeta}{\kappa(1-2\zeta)}\left(1 - \frac{1}{y}\right), \label{ode3}
\end{align}
and the constraint for the massless scalar field ((\ref{eqa}) with $V_0 w=0$) reads
\begin{equation} \label{eqaml}
 (\gamma+\kappa)^2 \frac{1-2\zeta}{1-\zeta} = 1 + \kappa^2 - \frac{1}{y}.
\end{equation}
Furthermore, regularity at the center (cf. (\ref{BV})) require
\begin{equation} \label{IV}
 w\rightarrow 0, \qquad y\rightarrow 1, \qquad \zeta\rightarrow 0, \qquad \gamma\rightarrow 0,
 \qquad \textrm{as} \qquad s\rightarrow -\infty \ \ (x\rightarrow 0),
\end{equation}
which give rise to the initial conditions. The objective is to solve the evolutionary equations (\ref{ode1}-\ref{ode3}) from $s=-\infty\ (u<0,r=0)$ to possibly $s=+\infty\ (u=0,r>0)$ (cf. Fig. \ref{fig4}). Among solutions subject to (\ref{IV}), black hole formation is indicated by integral curves reaching the plane $y=0$, where it locates an apparent horizon (cf. (\ref{mass})). Hence for naked singularities, we need to identify integral curves that can avoid contacting $y=0$ for the entire range of $s$. For this purpose, solutions terminating at a finite critical or singular point as $s\rightarrow +\infty$ are probable candidates.

In addition, the system (\ref{ode1}-\ref{ode3}) is invariant under the mapping $\kappa\rightarrow -\kappa$, $\gamma\rightarrow -\gamma$ with $y$ and $\zeta$ unchanged. Without loss of generality, we only consider $\kappa>0$ henceforward.

\section{Local behaviors} \label{sec:local}

Before we investigate global solutions, it is important to understand local phase space near critical and singular points (see Table \ref{table} for a quick summary). At these key locations, we can determine properties such as stability, dimensions of stable/unstable manifolds, uniqueness of solutions, etc., to help us build a fuller picture. Because the phase space for the massless scalar field is contained as a surface (signified by the constraint (\ref{eqaml})) in our 3-dimensional phase space, we expect that many discussions on the surface \cite{Christo94} can be transferable to the whole space.

\subsection{The initial point $s = -\infty$} \label{subsec:IP}

The initial point
\begin{equation}
 \mathcal{O}: \ \ y=1, \ \ \zeta=0, \ \ \gamma=0,
\end{equation}
is also a critical point of the vector field defined by (\ref{ode1}-\ref{ode3}). Therefore, we can determine the behavior of solutions in a neighborhood of $\mathcal{O}$ by the standard linearization method. With
\begin{equation}
 y = 1 + x_1, \qquad \zeta = x_2, \qquad \gamma = x_3,
\end{equation}
the Taylor series expansion in $x_{1,2,3}$ gives rise to a linearized form of (\ref{ode1}-\ref{ode3}):
\begin{equation} \label{lineq1}
 \dot{\mathbf{x}} = A\,\mathbf{x}, \qquad \mathbf{x} = (x_1,x_2,x_3)^T,
 \qquad A =
 \begin{pmatrix}
 0 & \kappa^2 & -2\kappa \\
 0 & 1 & 0 \\
 -\kappa^{-1} & 0 & 1
 \end{pmatrix}
 .
\end{equation}
The Jacobian matrix $A$ has eigenvalues
\begin{equation} \label{eigen_O}
 \lambda_1 = -1, \qquad \lambda_2 = 1, \qquad \lambda_3 = 2.
\end{equation}
Thus $\mathcal{O}$ is a saddle point. The general solution of (\ref{lineq1}) is given by
\begin{equation}
 x_1 = c_1 \rme^{\lambda_1 s} - c_3 \rme^{\lambda_3 s}, 
 \ x_2 = c_2 \rme^{\lambda_2 s},
 \ x_3 = \frac{c_1 \rme^{\lambda_1 s} + c_2 \kappa^2 \rme^{\lambda_2 s} + 2c_3 \rme^{\lambda_3 s}}{2\kappa},
\end{equation}
with three integration constants $c_{1,2,3}$. Now we recall the conditions (\ref{BV}), (\ref{wyzs}), and (\ref{zzeta}). They together imply
\begin{equation} \label{IV1}
 \lim_{s\rightarrow-\infty} \zeta\, \rme^{-s} = 2,
\end{equation}
and hence set $c_2=2$. Consequently, solutions subject to the conditions (\ref{IV}) and (\ref{IV1}) take the form
\begin{equation} \label{seriesL}
 y = 1 - c_3 \rme^{2s} + O(x_i x_j), 
 \ \zeta = 2\rme^s + O(x_i x_j),
 \ \gamma = \kappa \rme^s + \frac{c_3}{\kappa}\rme^{2s} + O(x_i x_j),
\end{equation}
which admits one free parameter $c_3$ associated with $\lambda_3$ and $i,j=1,2,3$. Particularly, the massless solution, which satisfies (\ref{eqaml}), occurs uniquely at $c_3=7\kappa^2/3$. The presence of a free parameter can also be confirmed by the Taylor series solution at $\mathcal{O}$ in terms of $x$. More specifically, we have
\begin{align}
 y &= 1 - c_T x^2 - 2c_T x^3 + O(x^4), \nonumber \\
 \zeta &= 2x - 4x^2 + (2c_T+8-\kappa^2)x^3 + O(x^4), \label{seriesT} \\
 \gamma &= \kappa x + \left(\frac{c_T}{\kappa}+\kappa\right)\! x^2
 + \left[\frac{3c_T}{\kappa} + (c_T+1)\kappa - \frac{\kappa^3}{2}\right]\! x^3 + O(x^4), \nonumber
\end{align}
with $c_T=\kappa^2/3$ for the unique massless solution. These series solutions set our first step toward global solutions. They will also be useful for generating approximate initial values near $\mathcal{O}$ for numerical integration.

Compared to Christodoulou's massless case, the interior solution ($0<\zeta<1/2$) is no longer unique. Instead, we have a 1-parameter family of integral curves emitting from $\mathcal{O}$. This enlarged solution manifold is due to the addition of the third eigenvalue $\lambda_3$ ($\lambda_{1,2}$ identical with the massless case \cite{Christo94}), which will lead to richer behavior.

\begin{figure}[htbp]
  \centering
  \mbox{
  \includegraphics[width=0.48\textwidth]{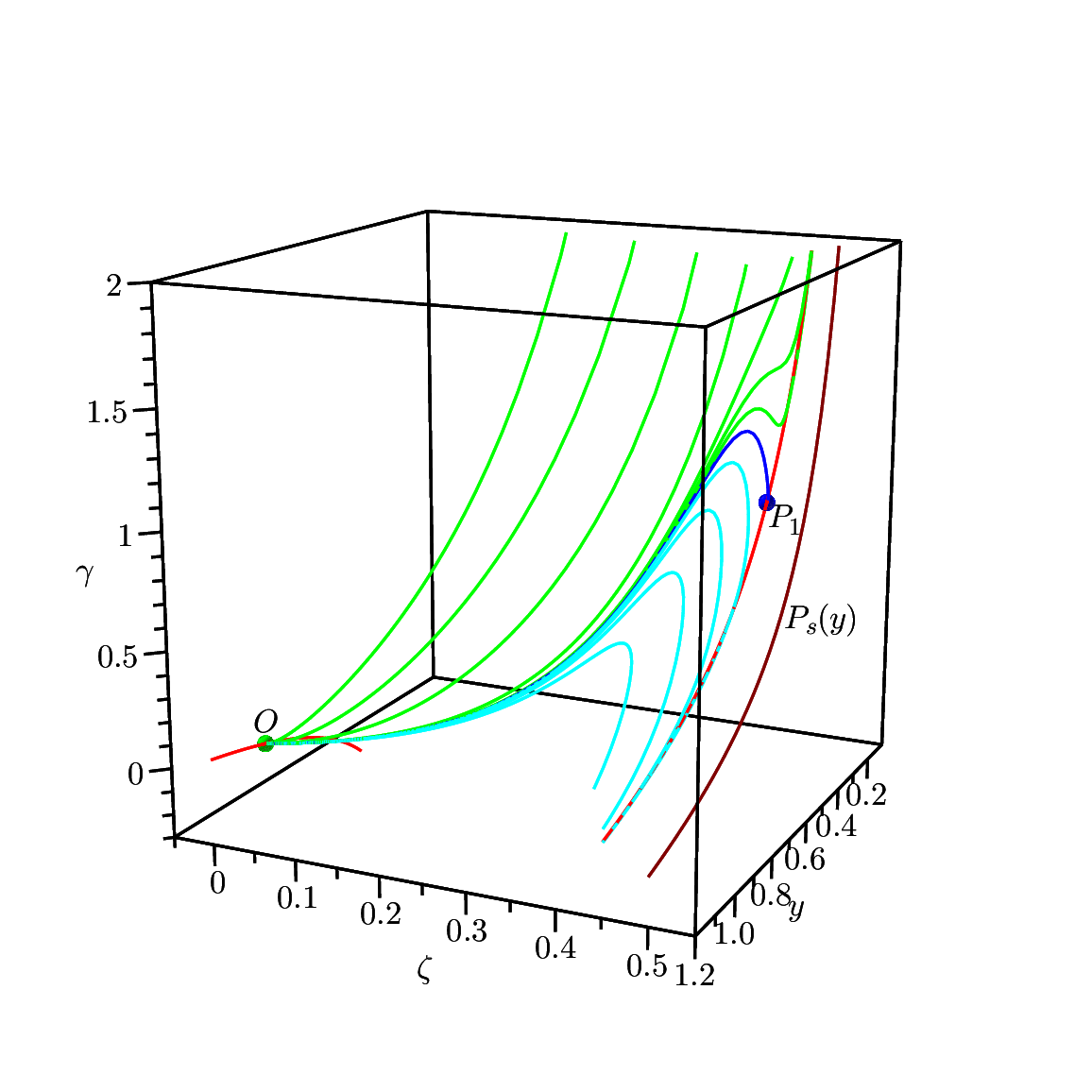}
  \includegraphics[width=0.48\textwidth]{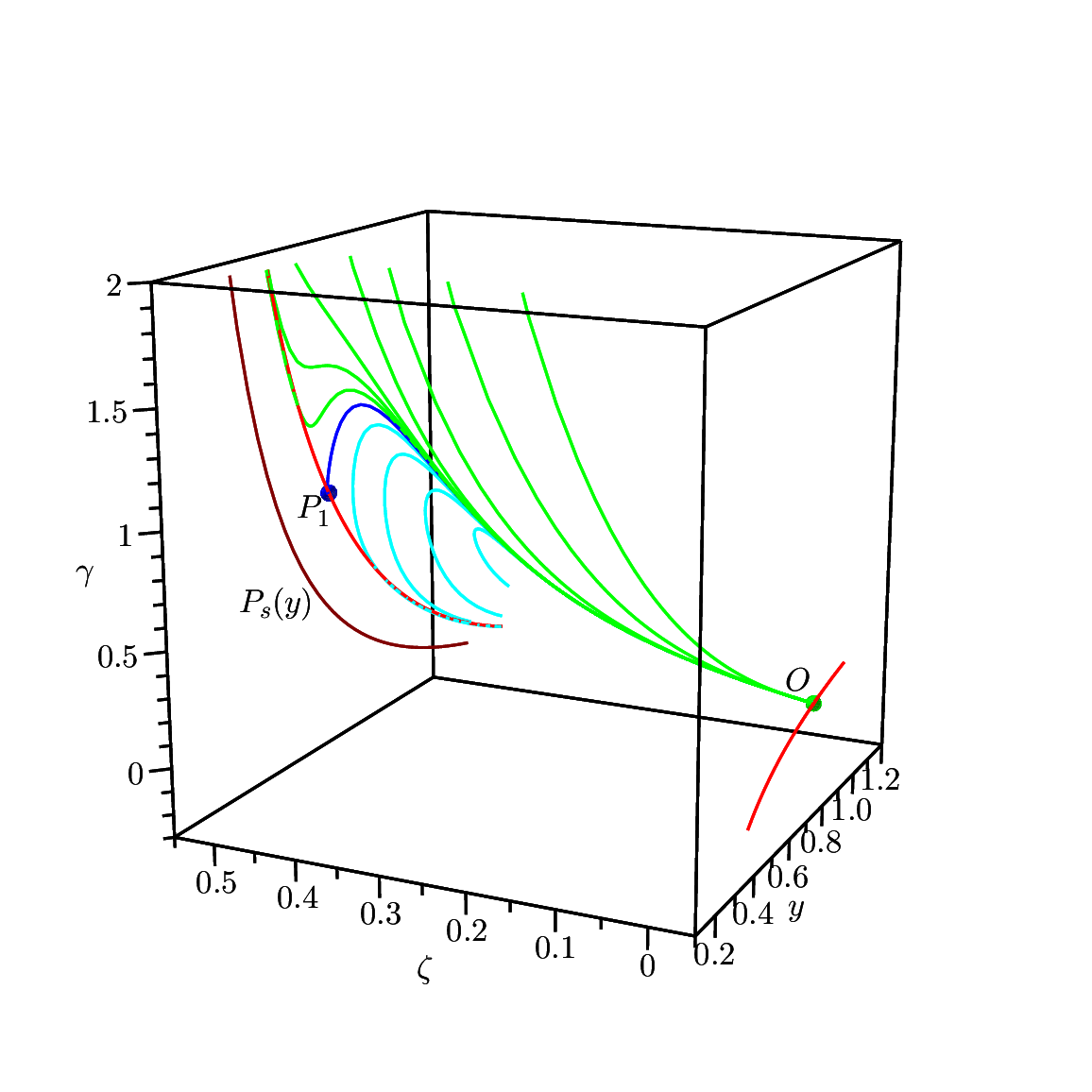}
  }
  \caption{Numerical integration with $\kappa=1.2$. The right plot shows the back view of the left. Green ($c_T>\kappa^2/3$, $V_0=1$, cf. (\ref{V0_cT})), blue, and cyan ($c_T<\kappa^2/3$, $V_0=-1$) curves emanating from $\mathcal{O}$ are members of the 1-parameter family of solutions of the system (\ref{ode1}-\ref{ode3}) subject to (\ref{IV}) and (\ref{IV1}). The negatively unstable manifold of $\mathcal{O}$ and the unstable manifold of $\mathcal{P}_1$ are marked by red curves. The blue curve ($c_T=\kappa^2/3$, $V_0=0$) ending at $\mathcal{P}_1$ delineates the unique massless solution. The maroon curve represents the set of singular points $\mathcal{P}_s(y)$ with $\zeta=1/2$. }
  \label{fig1}
\end{figure}

So far, we have not specified the sign of the scalar potential since the factor $V_0$ does not appear in (\ref{ode1}-\ref{ode3}). To determine the sign, we plug the series solution (\ref{seriesT}) into (\ref{eqa}) and obtain
\begin{equation}
 0 \leq \rme^{2\bar{h}/\kappa} = \frac{w}{x^2} = \frac{3c_T-\kappa^2}{2V_0} (1+2x) + O(x^2).
\end{equation}
For a real scalar field, the above quantity must be kept non-negative, which then sets
\begin{equation} \label{V0_cT}
 V_0 = \left\{
 \begin{array}{rl}
 -1 &\ \text{if } c_T < \kappa^2/3 \ \ (c_3 < 7\kappa^2/3), \\
 +1 &\ \text{if } c_T > \kappa^2/3 \ \ (c_3 > 7\kappa^2/3).
 \end{array} \right.
\end{equation}
Likewise, using (\ref{seriesL}), we have an equivalent sign choice in terms of $c_3$ given in the parentheses above. Therefore, according to the sign of $V_0$, the 2-dimensional interior solution manifold is divided into two parts by the unique massless solution ($c_T=\kappa^2/3$, $V_0=0$) (cf. Fig. \ref{fig1}).

\subsection{Critical endpoints $s=+\infty$} \label{sec:critpt}

Besides the initial point, the system possesses three more critical points that may serve as ending points of solution curves. We can treat them by linearization as well.

For the first critical point
\begin{equation}
 \mathcal{P}_1: \ \ y = \frac{1}{2},\ \ \zeta = \frac{2}{3+\kappa},\ \ \gamma = 1,
\end{equation}
the matrix $A$ reads
\begin{equation}
 A =
 \begin{pmatrix}
 0 & \frac{1}{2}(3+\kappa)^2 & -\kappa \\
 0 & -2(1+\kappa) & -\frac{4(1-\kappa)(1+\kappa)}{(3+\kappa)^2} \\
 \frac{4(1+\kappa)}{\kappa(1-\kappa)} & \frac{(1+\kappa+\kappa^2)(\kappa+3)^2}{\kappa(1-\kappa)}
 & \frac{2(1+\kappa-\kappa^3)}{\kappa(1-\kappa)}
 \end{pmatrix},
\end{equation}
with eigenvalues
\begin{equation} \label{eigen_P1}
 \lambda_1 = \frac{\kappa-\sqrt{4-3\kappa^2}}{1-\kappa},
 \qquad \lambda_2 = \frac{\kappa+\sqrt{4-3\kappa^2}}{1-\kappa},
 \qquad \lambda_3 = \frac{2(1+\kappa)}{\kappa}.
\end{equation}
Particularly, we have $\mathrm{Re}\lambda_{1,2}<0$ and $\lambda_3>0$ when $\kappa>1$, and $\lambda_1<0$, $\lambda_{2,3}>0$ when $0<\kappa<1$. Thus $\mathcal{P}_1$ is a saddle point. For the second critical point
\begin{equation}
 \mathcal{P}_{-1}: \ \ y = \frac{1}{2},\ \ \zeta = \frac{2}{3-\kappa},\ \ \gamma = -1,
\end{equation}
the matrix $A$ reads
\begin{equation}
 A =
 \begin{pmatrix}
 0 & \frac{1}{2}(3-\kappa)^2 & -\kappa \\
 0 & -2(1-\kappa) & \frac{4(1-\kappa)(1+\kappa)}{(3+\kappa)^2} \\
 \frac{4(1-\kappa)}{\kappa(1+\kappa)} & -\frac{(1-\kappa+\kappa^2)(\kappa-3)^2}{\kappa(1+\kappa)}
 & -\frac{2(1-\kappa+\kappa^3)}{\kappa(1+\kappa)}
 \end{pmatrix},
\end{equation}
with eigenvalues
\begin{equation} \label{eigen_P-1}
 \lambda_1 = \frac{-\kappa-\sqrt{4-3\kappa^2}}{1+\kappa},
 \ \lambda_2 = \frac{-\kappa+\sqrt{4-3\kappa^2}}{1+\kappa},
 \ \lambda_3 = -\frac{2(1-\kappa)}{\kappa}.
\end{equation}
Particularly, we have $\mathrm{Re}\lambda_{1,2}<0$ and $\lambda_3>0$ when $\kappa>1$, and $\lambda_{1,3}<0$, $\lambda_2>0$ when $0<\kappa<1$. Thus $\mathcal{P}_{-1}$ is also a saddle point.

Both of these points reside in the massless phase subspace determined by (\ref{eqaml}), and they are directly inherited from Christodoulou's case for having the same locations as well as the first two eigenvalues $\lambda_{1,2}$. Hence we have denoted them in the same way as in \cite{Christo94}.

The third critical point does not have a massless counterpart:
\begin{equation}
 \mathcal{P}_{c}: \ \ y = \frac{1}{1-\kappa^4},\ \ \zeta = 0,\ \ \gamma = -\kappa.
\end{equation}
However, as we will see in later sections, it is not of much relevance to our discussion. Here we only list the matrix $A$ and its eigenvalues below without further ado:
\begin{equation}
 A =
 \begin{pmatrix}
 0 & 0 & \frac{2\kappa}{1-\kappa^4} \\
 0 & 1-\kappa^2 & 0 \\
 -\frac{(1-\kappa^4)^2}{\kappa} & 0 & -1-\kappa^2
 \end{pmatrix},
\end{equation}
\begin{equation}
 \lambda_{1,2} = \frac{-1-\kappa^2 \mp \sqrt{(1+\kappa^2)(9\kappa^2-7)}}{2},
 \qquad \lambda_3 = 1-\kappa^2.
\end{equation}

\subsection{Singular endpoints $\zeta = 1$, $s=+\infty$} \label{sec:singlpt}

Besides the points $\mathcal{P}_{\pm 1}$ , Christodoulou also discussed a third critical point
\begin{equation} \label{P0}
 \mathcal{P}_0: \ \ y=\frac{1}{1+\kappa^2},\ \ \zeta = 1,\ \ \gamma = -\kappa,
\end{equation}
in the 2-dimensional phase space, which turned out to be crucial since solutions representing naked singularities tend to this point as $s\rightarrow +\infty$. For our 3-dimensional problem, $\mathcal{P}_0$ expands into a straight line along the $y$-direction:
\begin{equation} \label{P0y0}
 \mathcal{P}_0(y_0): \ \ y=y_0\neq 0,\ \ \zeta = 1,\ \ \gamma = -\kappa.
\end{equation}
Points on the line are critical for the equations (\ref{ode2}-\ref{ode3}), but singular for (\ref{ode1}).

To probe into the local behavior near $\mathcal{P}_0(y_0)$, we continue using
\begin{equation}
 y = y_0 + x_1, \qquad \zeta = 1+x_2, \qquad \gamma = -\kappa + x_3,
\end{equation}
as in previous subsections and perform (Laurent/power) series expansions in terms of $x_{1,2,3}$. Only keeping terms up to the first order, we arrive at
\begin{align}
 \dot x_1 &= 2 y_0 \kappa x_3 - y_0 \frac{x_3^2}{x_2}, \label{x1de} \\
 \dot x_2 &= (\kappa^2-1) x_2, \label{x2de} \\
 \dot x_3 &= \left(\frac{1-y_0}{y_0\kappa}+\kappa^3\right) x_2 - x_3. \label{x3de}
\end{align}
It is noted that a partial linearization succeeds in the $(\zeta,\gamma)$-plane since the linear equations (\ref{x2de}) and (\ref{x3de}) decouple from (\ref{x1de}) and do not depend on $x_1$. The matrix $A$ for $x_{2,3}$ alone reads
\begin{equation}
 \begin{pmatrix}
 \kappa^2-1 & 0 \\
 \frac{1-y_0}{y_0\kappa}+\kappa^3 & -1
 \end{pmatrix}.
\end{equation}
The eigenvalues are
\begin{equation} \label{eigen_P0}
 \lambda_1 = -1, \qquad \lambda_2 = -(1-\kappa^2),
\end{equation}
which coincide with the massless case (cf. \cite{Christo94}, (3.16)). Solving the linear equations gives
\begin{equation} \label{P0_zgamma}
 \zeta = 1 + c_2 \rme^{-(1-\kappa^2)s} + \cdots,
 \ \gamma = -\kappa + c_2 \left(\frac{1-y_0}{y_0\kappa^3}+\kappa\right)\rme^{-(1-\kappa^2)s}
 + c_1 \rme^{-s} + \cdots,
\end{equation}
with two parameters $c_{1,2}$. Fortunately, the troublesome term $x_3^2/x_2$ in (\ref{x1de}) is well-behaved under (\ref{P0_zgamma}), and we obtain formally
\begin{align}
 y = y_0 + \frac{c_2(1-y_0+y_0\kappa^4)(1-y_0-y_0\kappa^4)}{\kappa^6(1-\kappa^2) y_0} \rme^{-(1-\kappa^2)s} \nonumber \\
 + \frac{2c_1(1-y_0)}{\kappa^3} \rme^{-s}
 + \frac{y_0 c_1^2}{c_2(1+\kappa^2)} \rme^{-(1+\kappa^2)s} + \cdots, \label{P0_y}
\end{align}
where quadratic and higher-order terms in $x_{1,2,3}$ are omitted. Hence for $\kappa^2<1$, the asymptotic expansions indicate that each $\mathcal{P}_0(y_0)$ acts like an attractor (except in the $y$-direction). This will be later confirmed by numerical calculation. Very much like the massless solutions, $\mathcal{P}_0(y_0)$ will serve as the endpoints of naked singularity solutions (see Sec. \ref{sec:4D}).

\subsection{Singular (connecting) points $\zeta = \frac{1}{2}$, $s=s_*$}

If a continuous integral curve from $\mathcal{O}$ were to reach a point $\mathcal{P}_0(y_0)$ with $\zeta=1$, it would have to cross the plane $\zeta=1/2$ at the following set of singular points:
\begin{equation}
 \mathcal{P}_s(y_*): \ \ y = y_*\neq 0,\ \ \zeta = \frac{1}{2},\ \ \gamma = \frac{1-y_*}{y_*\kappa^3},
\end{equation}
which is imposed by (\ref{ode3}) (multiplied by $1-2\zeta$ on both sides). Hence if an interior solution attains $\mathcal{P}_s(y_*)$ at some finite $s=s_*$, it is necessary to extend it past $\zeta=1/2$.

To tackle these singular points where the uniqueness of solutions fails, we follow Christodoulou's method and introduce a new independent variable $t$ satisfying
\begin{equation} \label{ts}
 \frac{\rmd s}{\rmd t} = -\frac{1}{\zeta} + 2.
\end{equation}
In terms of $t$, the singular point $\mathcal{P}_{s}(y_*)$ becomes a critical point of the transformed equations. Then perform linearization with
\begin{equation}
 y = y_* + x_1(t), \qquad \zeta = \frac{1}{2} + x_2(t), \qquad \gamma = \frac{1-y_*}{y_*\kappa^3} + x_3(t),
\end{equation}
and calculate the matrix $A$ as
\begin{equation}
 \begin{pmatrix}
 0 & \frac{4(-1+y_*+y_*\kappa^4)(1-y_*+y_*\kappa^4)}{y_*\kappa^6} & 0 \\
 0 & 1-\kappa^2 & 0 \\
 \frac{1}{y_*^2\kappa} & \frac{4(1-y_*)(1-y_*+y_*\kappa^4)(1-y_*+y_*\kappa^2+y_*\kappa^4)}{y_*^3\kappa^9}
 & \kappa^2
 \end{pmatrix},
\end{equation}
along with its eigenvalues
\begin{equation} \label{eigen_Ps}
 \lambda_1 = \kappa^2, \qquad \lambda_2 = 1-\kappa^2, \qquad \lambda_3 = 0,
\end{equation}
which are independent of $y_*$. The zero eigenvalue $\lambda_3$ reflects the fact that the set of $\mathcal{P}_s(y_*)$ forms a curve, while the other two eigenvalues are identical to the massless case (cf. \cite{Christo94}, p. 620). Therefore, if $\kappa^2<1$, both $\lambda_{1,2}$ are positive and one can construct a 1-parameter family of solutions which approach $\mathcal{P}_s(y_*)$ as $t\rightarrow -\infty$ (or $s\rightarrow s_*$ from either side of $\zeta=1/2$) \cite{Christo94}. Consequently, a single interior solution curve from $\zeta<1/2$ may branch out at $\mathcal{P}_s(y_*)$ into a 2-dimensional solution manifold for $s>s_*$ (cf. Figs. \ref{fig2}, \ref{fig3}, and \ref{fig5}).

As an extra comment, we mention that the massless phase subspace determined by (\ref{eqaml}) only intersects the curve $\mathcal{P}_s(y)$ at one point:
\begin{equation}
 \mathcal{P}_{sc}:\ \ y = \frac{1}{1+\kappa^2}, \ \ \zeta = \frac{1}{2}, \ \ \gamma = \frac{1}{\kappa}.
\end{equation}
which was closely examined by Christodoulou \cite{Christo94}. In Sec. \ref{case2}, we will compare continuations of the solutions from $\mathcal{P}_{sc}$ with those from other points on $\mathcal{P}_s(y)$.

\subsection{Apparent horizon $y=0$}

The apparent horizon is signaled by $y=0$, which also renders the equation (\ref{ode3}) singular. To capture the dominant behavior near $y=0$, we perform the Painlev\'{e} analysis \cite{Conte08} and identify the following Puiseux series expansions:
\begin{align}
 y &= a_1\sqrt{s_\infty-s} - \left[\frac{a_1\kappa^2}{3a_2} + 4a_1a_2
 \pm \left(\frac{4\sqrt{2}a_1\kappa}{3} + \frac{\sqrt{2}(2a_1a_2-1)}{3a_2\kappa} \right)\right](s_\infty-s)
 + \cdots, \label{yAH} \\
 \zeta &= \frac{1}{2} + a_2\sqrt{s_\infty-s} + \cdots, \\
 \gamma &= \pm \frac{1}{\sqrt{2} \sqrt{s_\infty-s}} - \frac{2\kappa}{3} - \frac{2a_1a_2-1}{6a_1a_2\kappa}
 \mp \frac{\sqrt{2}\kappa^2}{12a_2} + \cdots, \label{gammaAH}
\end{align}
which converges to
\begin{equation}
 \mathcal{P}_{AH}=\mathcal{P}_s(0):\ \ y = 0,\ \ \zeta = \frac{1}{2},\ \ \gamma = \pm \infty,
\end{equation}
as $s\rightarrow s_\infty-$. This series solution is general for having three free parameters $a_1$, $a_2$ and $s_\infty$. Thereby, one may regard the apparent horizon $\mathcal{P}_{AH}$ effectively as an \emph{attractor} in the phase space. The leading order terms of the expansions agree with Christodoulou's Proposition 3.2 in \cite{Christo94}.

Using (\ref{gammaAH}), we can estimate the scalar field near the apparent horizon as
\begin{equation}
 \bar{h}(x) = \int \frac{\gamma(x)}{x} \rmd x
 \simeq \frac{1}{\sqrt{2}} \int \frac{1}{x \sqrt{\ln x_{AH} - \ln x}} \rmd x
 = -\sqrt{2 \ln\!\left(\frac{x_{AH}}{x}\right)} + \bar{h}(x_{AH}),
\end{equation}
with $\ln x_{AH} = s_\infty$ and $x\leq x_{AH}$. Hence, the scalar field maintains finite and continuous at the apparent horizon and may be further extended to infinity.

%
\begin{table}
  \caption{\label{table} Summary of important points in the phase space. }
  \begin{tabular}{ccccc}
  Pts & $(y,\zeta,\gamma)$ & Description & Stability & Eigenvalues \\ \hline
  $\mathcal{O}$ & $(1,0,0)$ & initial pt $s=-\infty\ (u<0, r=0)$ & saddle & (\ref{eigen_O}) \\
  $\mathcal{P}_1$ & $(\frac{1}{2},\frac{2}{3+\kappa},1)$
  & critical endpt $s=+\infty\ (u=0, r>0)$ & saddle & (\ref{eigen_P1}) \\
  $\mathcal{P}_{-1}$ & $(\frac{1}{2},\frac{2}{3-\kappa},-1)$
  & critical endpt $s=+\infty$ & saddle & (\ref{eigen_P-1}) \\
  $\mathcal{P}_0(y)$ & $(y\neq 0,1,-\kappa)$
  & singular endpts $s=+\infty$, $\kappa\in (0,1)$ & attracting & (\ref{eigen_P0}) \\
  $\mathcal{P}_s(y)$ & $(y\neq 0,\frac{1}{2},\frac{1-y}{y\kappa^3})$
  & singular pts $s=s_*$, $\kappa\in (0,1)$ &  & (\ref{eigen_Ps}) \\
  $\mathcal{P}_{sc}$ & $(\frac{1}{1+\kappa^2},\frac{1}{2},\frac{1}{\kappa})$
  & singular pt $s=s_*$, $\kappa\in (0,1)$ &  & (\ref{eigen_Ps}) \\
  $\mathcal{P}_{AH}$ & $(0,\frac{1}{2},\pm\infty)$
  & apparent horizon $s=s_\infty$ & attracting & \\
  \end{tabular}
\end{table}

\section{Global solutions} \label{sec:global}

We move on to constructing global solutions by numerical integration and verifying our previous analysis of local behaviors. To systematically create numerical curves from the initial (stationary) point $\mathcal{O}$, we use the truncated series solution (\ref{seriesL}) or (\ref{seriesT}) to generate initial values near $\mathcal{O}$ (e.g., taking $s=-10$ and $c_3=1,2,\cdots$), and then integrate both backward and forward with the standard fourth-fifth order Runge-Kutta method. This simplifies numerical procedures and fulfills the conditions (\ref{IV}) and (\ref{IV1}) to a satisfactory accuracy. Moreover, upon nearing the singular curve $\mathcal{P}_s(y)$ or the plane $y=0$, it is also desirable that step sizes of integration be kept small so as to produce more accurate plotting. Similar procedures also apply to integral curves drawn from other critical or singular points.

As we have seen in local studies, the nature of the critical and singular points relies heavily on the value of the parameter $\kappa$. Our numerical results split into two cases.

\subsection{Case $\kappa > 1$} \label{sec:case1}

In our first set of plots with $\kappa=1.2$, solutions of interest are limited to $0<\zeta<1/2$ with $\mathcal{P}_1$ involved. Since $\mathcal{P}_1$ is a saddle point, the 1-dimensional unstable manifold (red curve) can draw away incoming integral curves (green and cyan) from $\mathcal{O}$ on either side of the transverse 2-dimensional stable manifold (cf. Sec. \ref{sec:critpt}). This naturally classifies the solutions into three types. For solutions heading toward the plane $y=0$ (green curves), they have $c_T>\kappa^2/3$ and correspond to $V_0=1$ according to (\ref{V0_cT}). From the series expansion (\ref{yAH}-\ref{gammaAH}), one has $y\rightarrow 0$ and $\gamma\rightarrow \infty$ as $s\rightarrow s_\infty-$ for some finite $s_\infty$. Numerical integration of the green curves is consistent with this limiting behavior in that the curves quickly exceed the maximum representable numerical range (for $\gamma$) of the computer program within a finite $s$ while ascending in the direction of $\mathcal{P}_{AH}$. Hence the first type represents black hole formation. On the other side of the stable manifold, solutions moving away from $y=0$ (cyan curves) have $V_0=-1$, and their Misner-Sharp masses defined in (\ref{mass}) can assume large negative values. The intermediate between the two types is an exceptional solution (blue curve) terminating at $\mathcal{P}_1$ as $s\rightarrow+\infty$. It also lies in the massless phase subspace determined by (\ref{eqaml}), hence corresponding to the unique interior solution by Christodoulou \cite{Christo94}. According to \cite{Brady95}, the associated spacetime has a central singularity at $u=0$ that is not visible to observers at a finite radius.

Unlike the massless evolution, black hole formation completely takes over under positive exponential potentials. This effect appears in line with cosmic censorship, which will be further discussed in later sections. For other values of $\kappa>1$, the qualitative behavior we have described remains the same.

\subsection{Case $0<\kappa < 1$} \label{case2}

\begin{figure}[htbp]
  \centering
  \mbox{
  \includegraphics[width=0.45\textwidth]{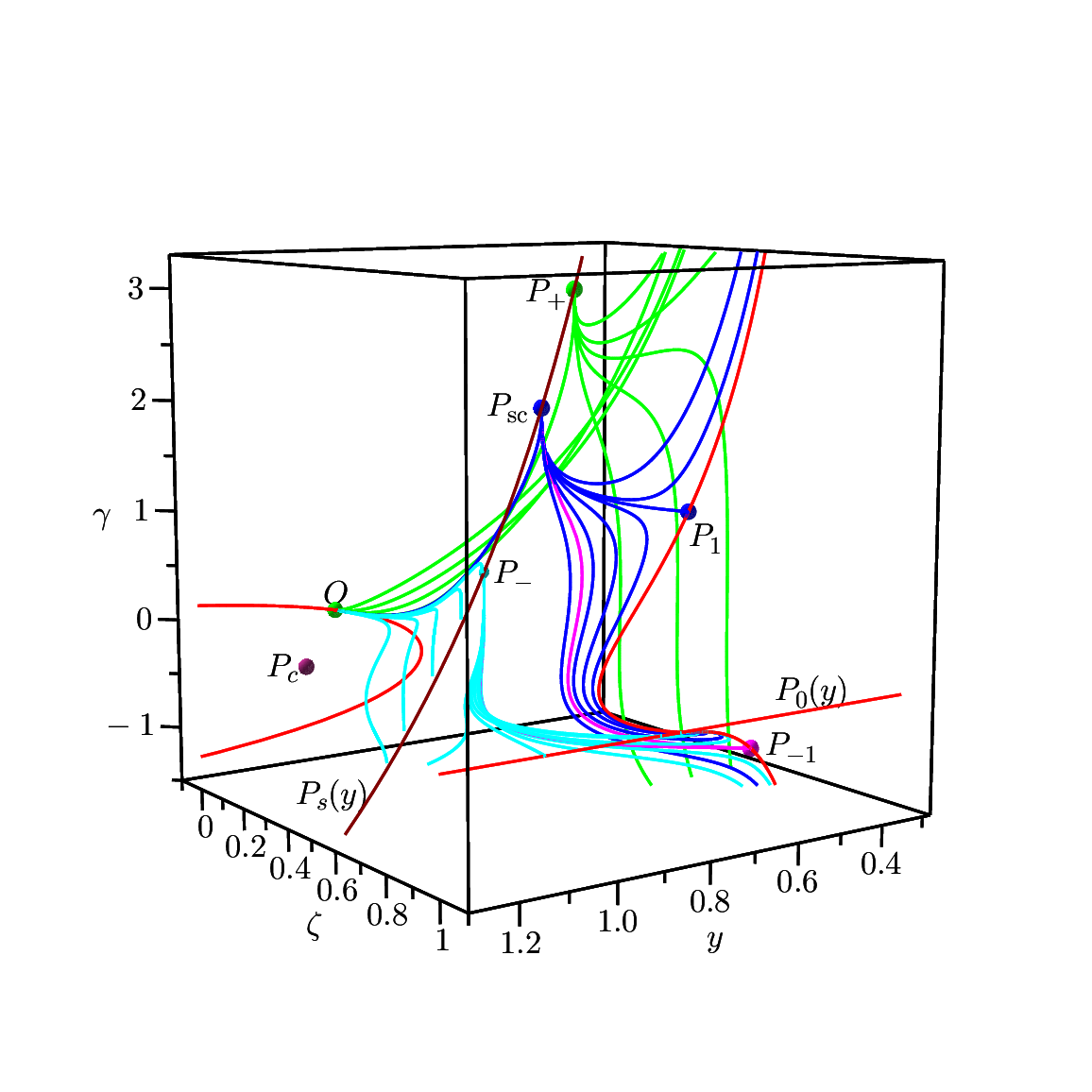}
  \includegraphics[width=0.45\textwidth]{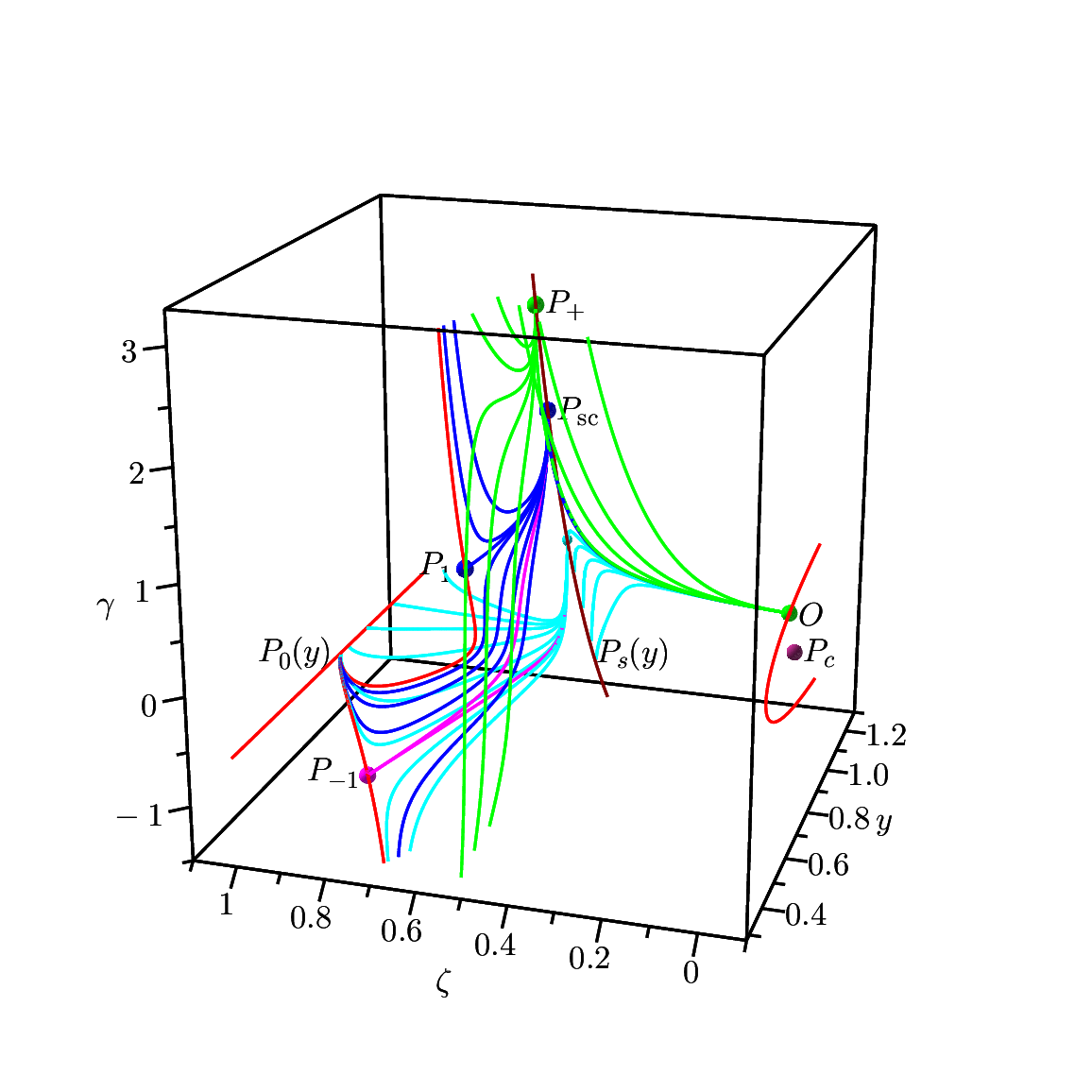}
  }
  \includegraphics[width=0.5\textwidth]{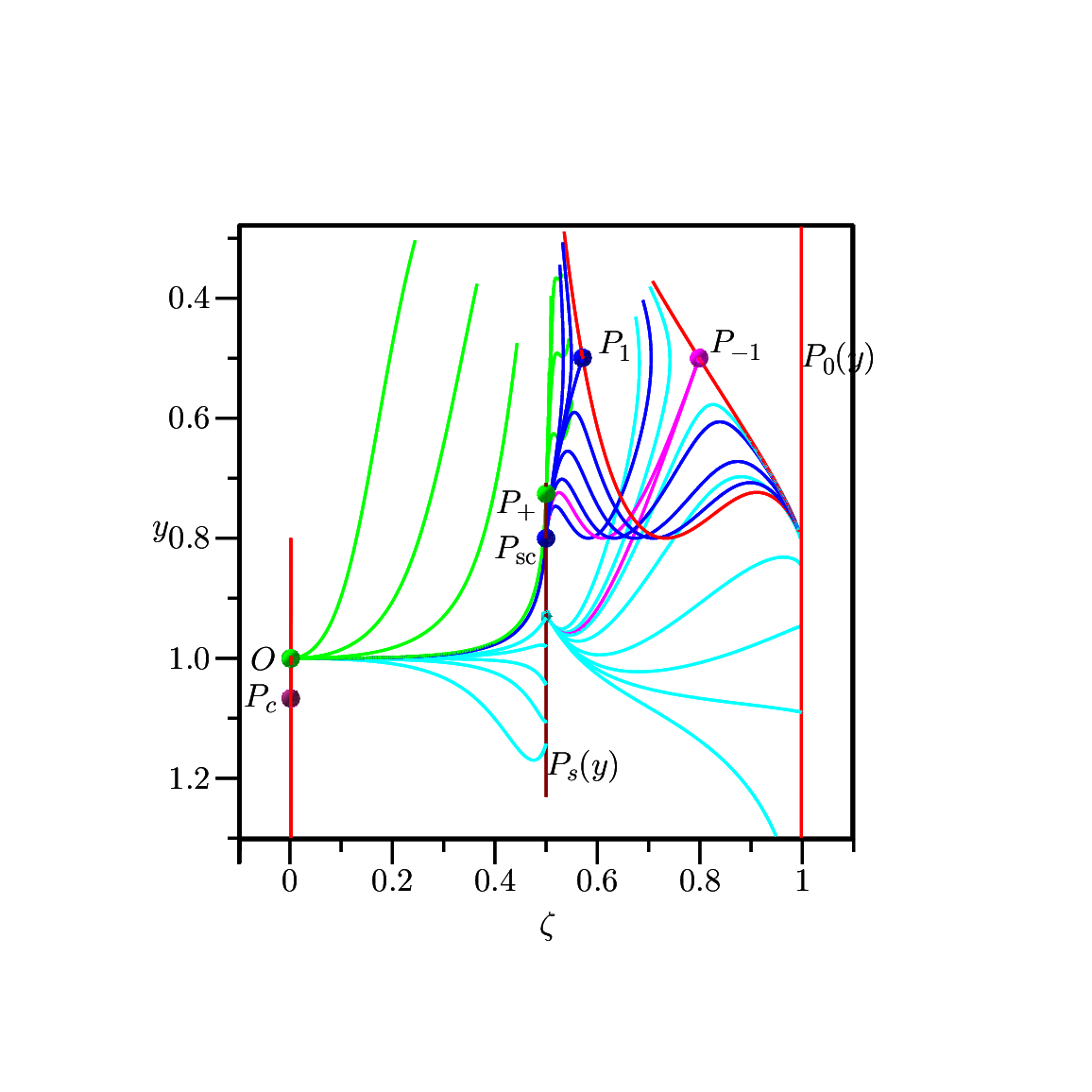}
  \caption{Numerical integration with $\kappa=0.5$ (roughly the front and back views, and the projection onto the $y$-$\zeta$ plane). Integral curves from $\mathcal{O}$ may reach the singular curve $\mathcal{P}_s(y)$ (maroon). Particularly, the blue ($V_0=0$) curve attaining $\mathcal{P}_{sc}$ is the unique massless interior solution interpolating between green ($V_0=1$) and cyan ($V_0=-1$) curves. In the region $\zeta>1/2$, curves drawn from $\mathcal{P}_{\pm}$ and $\mathcal{P}_{sc}$ are continuations of the interior solutions. The two purple curves terminating at $\mathcal{P}_{-1}$ separate naked singularity solutions and black hole solutions. The red curve passing through $\mathcal{P}_1$ ($\mathcal{P}_{-1}$) with one end reaching $\mathcal{P}_0(y)$ represents the unstable manifold of $\mathcal{P}_1$ ($\mathcal{P}_{-1}$) in the massless phase subspace (\ref{eqaml}).
  }
  \label{fig2}
\end{figure}

\begin{figure}[htbp]
  \centering
  \mbox{
  \includegraphics[width=0.45\textwidth]{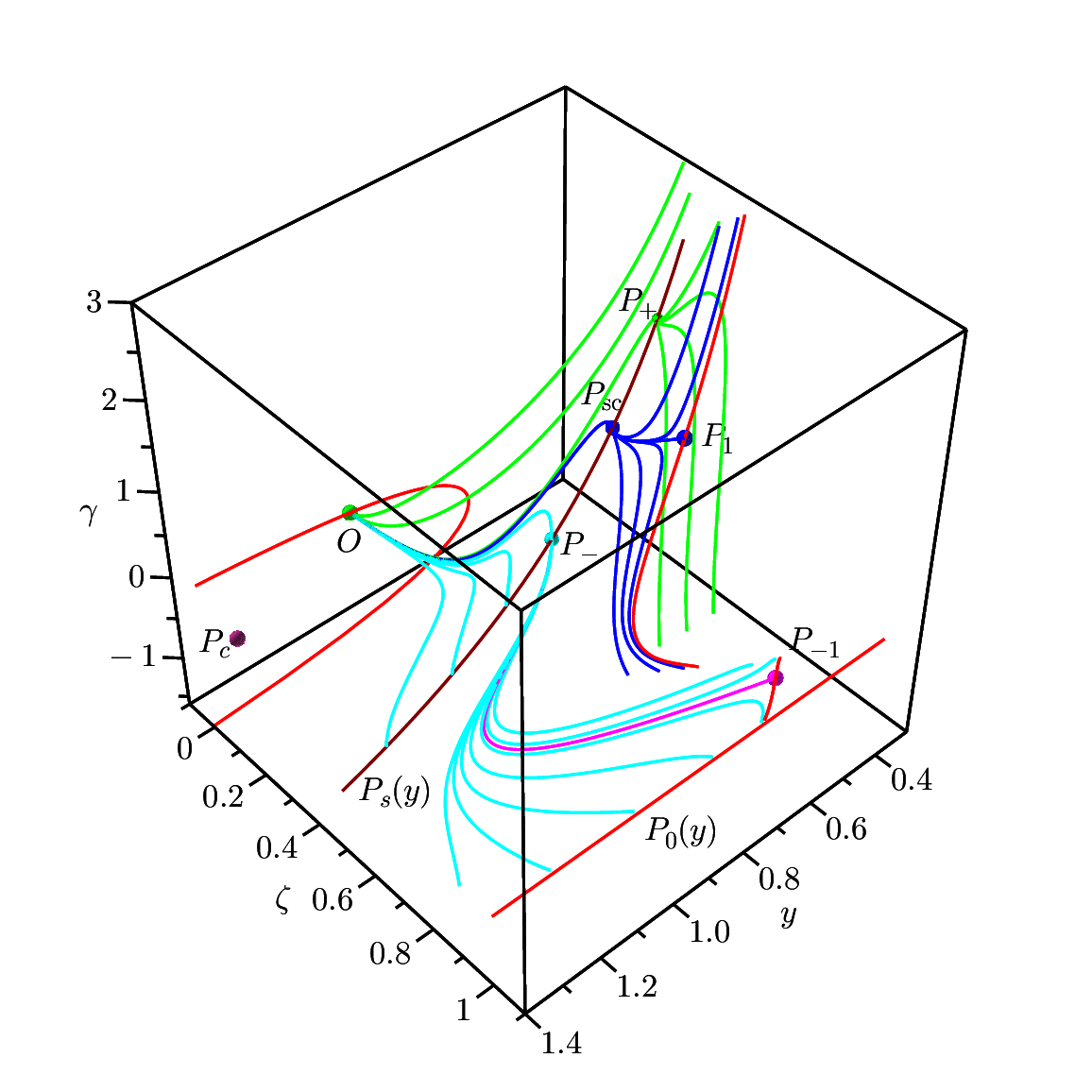}
  \includegraphics[width=0.45\textwidth]{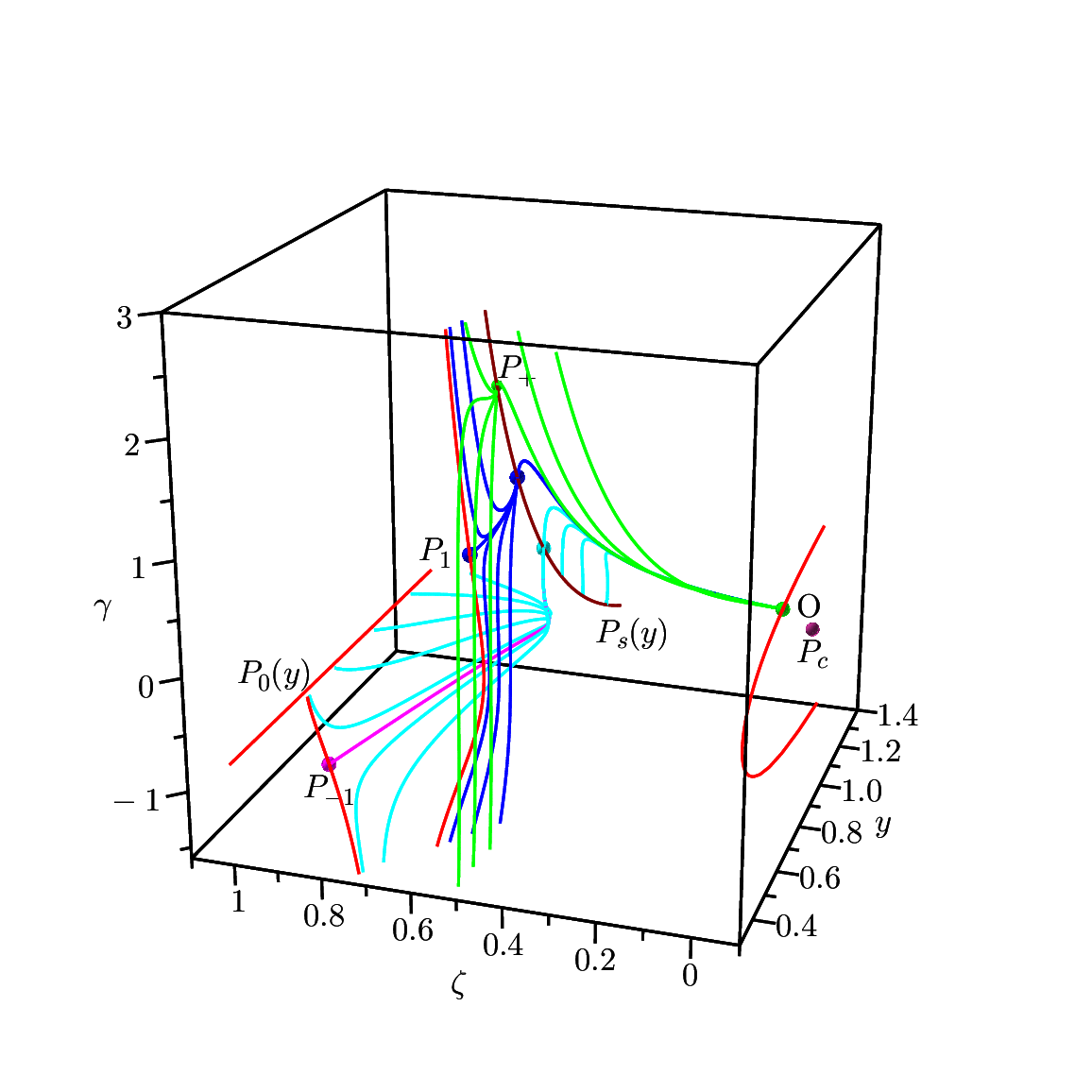}
  }
  \includegraphics[width=0.5\textwidth]{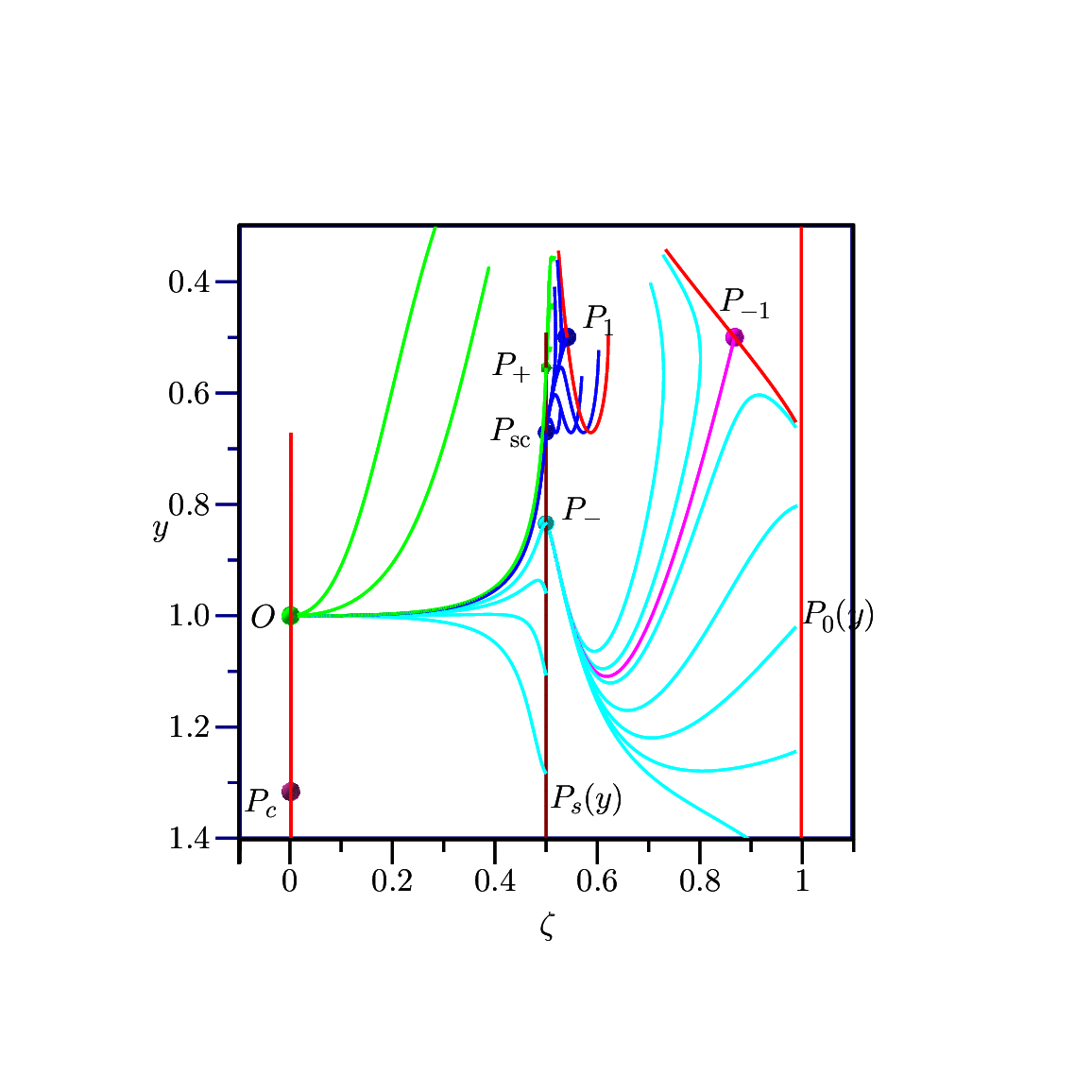}
  \caption{Numerical integration with $\kappa=0.7$ (roughly the upper front and back views, and the projection onto the $y$-$\zeta$ plane). Behavior of the green ($V_0=1$) and cyan ($V_0=-1$) curves are similar to those in Fig. \ref{fig2}. However, for massless solutions, the blue curves and the unstable manifold of $\mathcal{P}_1$ (red curve) no longer extend to the singular line $\mathcal{P}_0(y)$, indicating that naked singularities cease to occur.}
  \label{fig3}
\end{figure}

This case is much more involved as integral curves can pass through the plane $\zeta=1/2$. In the region $0<\zeta<1/2$, an interior solution curve shown in Fig. \ref{fig2} and Fig. \ref{fig3} can either move toward the apparent horizon $y=0$ or run into the singular curve $\mathcal{P}_s(y)$ in a finite $s$. In Sec. \ref{sec:singlpt}, we point out that once running into a point on $\mathcal{P}_s(y)$, an interior solution may branch into a one-parameter family of exterior solutions. This is illustrated in Fig. \ref{fig2} and Fig. \ref{fig3}, where we have picked out three representative locations on $\mathcal{P}_s(y)$---$\mathcal{P}_{sc}$ for $V_0=0$ and $\mathcal{P}_{\pm}$ for $V_0=\pm 1$---from which continuations are plotted. Amongst these prolonged curves, naked singularity solutions correspond to those approaching $\mathcal{P}_0(y)$ as $s\rightarrow +\infty$ (see Sec. \ref{sec:4D}). Clearly in Fig. \ref{fig2} with $\kappa=0.5$, such curves can emerge from $\mathcal{P}_{sc}$ (also $\mathcal{P}_-$, but not from $\mathcal{P}_+$). However, when $\kappa=0.7$ as in Fig. \ref{fig3}, they cease to show up in massless solutions altogether (blue curves). This numerical result agrees with Christodoulou's rigorous argument that, in the massless scalar field, naked singularities only exist when $\kappa^2<1/3$. To understand this phenomenon more intuitively from our plots, we note that the unstable manifold $\mathcal{M}_1$ (red curve, restricted on the surface (\ref{eqaml})) of the saddle point $\mathcal{P}_1$ appears to be a good indicator since it may dictate asymptotic behavior of the integral curves in its vicinity. More specifically in Fig. \ref{fig2}, one arm of $\mathcal{M}_1$ extends to $\mathcal{P}_0(y)$, thereby pulling solution curves nearby to the same endpoint. While in Fig. \ref{fig3}, both ends of $\mathcal{M}_1$ lead away from $\mathcal{P}_0(y)$, and are followed by other integral curves. Moreover, indicated by numerical computation, the critical case of $\kappa=1/\sqrt{3}$ has the lower segment of $\mathcal{M}_1$ falling exactly on the point $\mathcal{P}_{-1}$ \cite{Christo94}. No naked singularity arises in this critical case.

The situation, however, differs greatly for solutions with negative exponential potentials (cyan curves). Starting from a singular connecting point lower than $\mathcal{P}_{sc}$, an exterior solution may evade the influence of $\mathcal{P}_1$ and still reach out to $\mathcal{P}_0(y)$ regardless of $k^2<1/3$ or $1/3\leq k^2<1$. Besides naked singularities, other outcomes are also presented. In both Fig. \ref{fig2} and Fig. \ref{fig3}, by tuning the second parameter that controls the bifurcation, one can observe a transition between naked singularity and black hole solutions. The borderline separating these two extremes is demarcated by the critical solutions terminating at $\mathcal{P}_{-1}$ (purple curves; more generally, the stable manifold of $\mathcal{P}_{-1}$). The critical spacetime can be thought as sitting on the threshold of containing a naked singularity or a black hole, but just avoiding both. Similar behavior is also found in massless solutions (blue curves near $\mathcal{P}_1$ and $\mathcal{P}_{-1}$ in Fig. \ref{fig2}). As an aside, see \cite{Zhang15a,Zhang15b} for more examples of naked singularity/black hole transitions with non-trivial scalar potentials.

As regards solutions with $V_0=1$, black hole formation appears to be the only possible outcome since all green curves are heading to $\mathcal{P}_{AH}$ instead of $\mathcal{P}_0(y)$. This observation is further supported by numerical tests at other locations on $\mathcal{P}_s(y)$ above $\mathcal{P}_{sc}$. Therefore, we speculate that the naked singularity solutions by Christodoulou defy extension to positive exponential potentials.

We summarize our numerical results as follows: \\
\centerline{
\begin{minipage}[c][2.5cm]{0.95\textwidth}
 \emph{Naked singularity solutions, represented by integral curves that run continuously from $\mathcal{O}$---via $\mathcal{P}_s$---to $\mathcal{P}_0$, exist when}
 \begin{align}
  & 0<\kappa^2<1/3, \ \textit{if}\ \ V_0=0\ \text{\cite{Christo94}}, \label{kNS} \\
  & 0<\kappa^2<1, \ \textit{if}\ \ V_0=-1. \label{kNSV}
 \end{align}
\end{minipage}}

\section{Naked singularity formation in 4 dimensions} \label{sec:4D}

In this section, we examine various aspects of the 4-dimensional metric (\ref{metricx}) with inputs from the solution curves we have computed. The metric $\rmd s^2$  deserves attention in its own right since it directly builds upon Christodoulou's results.

To begin with, we show that for solutions terminating at $\mathcal{P}_0(y)$, the corresponding 4-dimensional spacetime indeed contains a curvature singularity at the point $u=0$, $r=0$ that is visible to distant observers (cf. Fig. \ref{fig4}). To see this, one can first calculate the Ricci scalar as
\begin{align}
 R &= - 2 T = 2 \big(\nabla_\alpha \phi \nabla^\alpha \phi + 4V(\phi) \big) \nonumber \\
 &= 2 \left[ -\frac{2}{g} \phi_{,u}\phi_{,r} + \frac{\tilde{g}}{g}(\phi_{,r})^2 + 4V_0 \exp\!\left(\frac{2\phi}{\kappa}\right)\right] \nonumber \\
 &= \frac{2\tilde{g}}{u^2 x^2 g} \left[\gamma^2 - \frac{2x}{\tilde{g}} (\gamma^2+\kappa\gamma)\right] + \frac{8V_0 w}{u^2x^2} \nonumber \\
 &= \frac{2y}{u^2 x^2} \left[\gamma^2 - \frac{\zeta}{1-\zeta}(\gamma^2+\kappa\gamma)\right] + \frac{8V_0 w}{u^2x^2}. \label{Ricci}
\end{align}
Then by applying the limit (cf. (\ref{P0_zgamma}))
\begin{align}
 y \rightarrow y_0 \geq \frac{1}{1+\kappa^2},
 \qquad (\zeta-1) \rme^{(1-\kappa^2)s} \rightarrow c_2<0, \nonumber \\
 (\gamma + \kappa) \rme^{(1-\kappa^2)s} \rightarrow c_2 \left(\frac{1-y_0}{y_0\kappa^3}+\kappa\right),
 \ \ \text{as}\ \ s\rightarrow +\infty \ (u\rightarrow 0-), \label{yzgP0}
\end{align}
one arrives at
\begin{equation}
 R = \frac{1}{r^2} \left[2(y_0-1) \left(\frac{1}{\kappa^2} - 2\right) - 4y_0\kappa^2\right],
\end{equation}
with $y_0=1/(1+\kappa^2)$ for the massless solutions and $c_2<0$ from numerical calculation. Clearly, this scalar invariant blows up at $r=0$, but remains finite elsewhere on the future light cone $u=0$ of the center (the Cauchy horizon). Moreover, one can verify that $R$ is also finite for $u<0$ given the regularity condition at the center and the expression (\ref{Ricci}) being continuous at $\mathcal{P}_s(y_*)$.

Now it remains to establish that the central curvature singularity is indeed visible, i.e., that the outgoing null geodesics $u=0$ can escape from $r=0$. Following the method in \cite{Christo94} (see also \cite{Bedjaoui10}), we investigate incoming radial null geodesics, assuming $\zeta>1/2$, and show that they can approach a finite $r$ as $u\rightarrow 0$. In terms of the variables $t=-\ln(-u)$ and $s=\ln(-r/u)$, the null geodesic equation
\begin{equation}
 \frac{\rmd r}{\rmd u} = -\frac{\tilde{g}}{2}
\end{equation}
can be rewritten as
\begin{equation}
 \frac{\rmd s}{\rmd t} = -\frac{1}{\zeta(s)} + 2 > 0.
\end{equation}
Integrating along a null ray originated at some $(t_0, s_0>s_*)$, we have
\begin{equation}
 t - t_0 = \int^s_{s_0} \frac{\zeta}{2\zeta-1} \rmd s'.
\end{equation}
The asymptotic series (\ref{P0_zgamma}) implies
\begin{equation}
 \left(\frac{\zeta}{2\zeta-1} - 1\right) \rme^{(1-\kappa^2)s} \rightarrow -c_2,
 \ \ \text{as}\ \ s\rightarrow +\infty.
\end{equation}
Therefore, the quantity
\begin{equation}
 \ln r = s - t = s_0 - t_0 - \int^s_{s_0} \left(\frac{\zeta}{2\zeta-1} - 1\right) \rmd s'
\end{equation}
converges to a finite limit as $t\rightarrow +\infty$ since the integral above is bounded. Also because $c_2<0$, we conclude that along an incoming null ray near $u=0$, the radius $r$ decreases to a finite value as $u$ increases to $0$.

Using the limit $y\rightarrow y_0$ as $s\rightarrow+\infty$, we can further determine the mass function at $u=0$ as
\begin{equation}
 m = \frac{1-y_0}{2}\, r,
\end{equation}
with $y_0=1/(1+\kappa^2)$ for the massless solutions and $y_0>1/(1+\kappa^2)$ for $V_0=-1$. Thus the central singularity itself has vanishing mass. Similarly, we can write down the scalar field in the limit $u\rightarrow 0$ as
\begin{equation} \label{phi}
 \phi = \frac{\kappa}{2}\ln\!\left(\frac{1-(1+\kappa^2)y_0}{2V_0}\right) - \kappa \ln r,
 \ V_0=-1, \ y_0>1/(1+\kappa^2),
\end{equation}
and for the massless scalar field,
\begin{equation}
 \phi = -\kappa \ln r,
\end{equation}
both of which diverge logarithmically at $r=0$. From the equation (\ref{phi}), the negative potential energy at $u=0$ obeys
\begin{equation}
 V \propto -\frac{1}{r^2},
\end{equation}
which is non-negligible near the center.

We comment that the 4-dimensional naked singularity solutions that we have considered are, by themselves, neither asymptotically flat nor AdS. This unusual asymptotic behavior is in fact typical for negative exponential potentials \cite{Poletti94,Chan95,Cai04}. Nevertheless, we should emphasize that since the past light cone of the singularity is given by $-r/u = \exp(s_*)$, the naked singularity formation is essentially a local event near the symmetry center and independent of the geometry far from it.

Finally we summarize the results on the 4-dimensional metric (\ref{metricx}). In the class of self-similar spacetimes, we have studied the dynamical system (\ref{ode1}-\ref{ode3}) and examined the effect of exponential potentials on the naked singularity solutions that were first constructed by Christodoulou and numerically explored by Brady. For every fixed $\kappa$, the original 2-dimensional phase space is embedded into the 3-dimensional phase space, which showcases more dynamical events. Using combined analytic and numerical methods, we build a pictorial representation of Christodoulou's solutions and put them in a richer context with new solutions. In our 3-dimensional plots, two competing ``attractors''---one for naked singularities and one for black holes---play a major role as time evolutions may be drawn into either one of them. In addition, transitions between them are identified. Our plots also evince that Christodoulou's solution manifold serves as a boundary that separates solutions with $V_0=1$ and $V_0=-1$. On one side of his solution manifold, a \emph{two}-parameter family of naked singularity solutions continues to exist under the negative potentials $V=-\exp(2\phi/\kappa)$ for $0<\kappa^2<1$. In contrast, on the side with positive exponential potentials, all numerical evolutions tested by us encounter apparent horizons, which suggests a strong censoring effect of the positive potentials. A similar situation is also found in the regime $\kappa^2>1$ (see Sec. \ref{sec:case1}).

\section{Naked singularity formation in 6 dimensions} \label{sec:6D}

\begin{figure}[htbp]
  \includegraphics[width=0.4\textwidth]{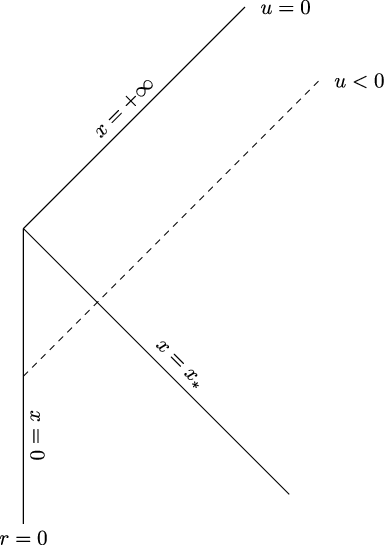}
  \caption{A spacetime diagram of the naked singularity formation near the symmetry center. The self-similar spacetime is obtained by solving the autonomous system (\ref{ode1}-\ref{ode3}) from $x=0$ to $x=+\infty$ (the coordinate line $x=-r/u$ sweeping from $r=0$ to $u=0$). The line $x=x_*$ ($x=+\infty$) marks the past (future) light cones of the central curvature singularity at $r=0$, $u=0$. There is no apparent horizon. The curvature invariants (the Ricci and Kretschmann scalars) are well-behaved for $u<0$ at the center and beyond. }
  \label{fig4}
\end{figure}

\begin{figure}[htbp]
  \includegraphics[width=0.5\textwidth]{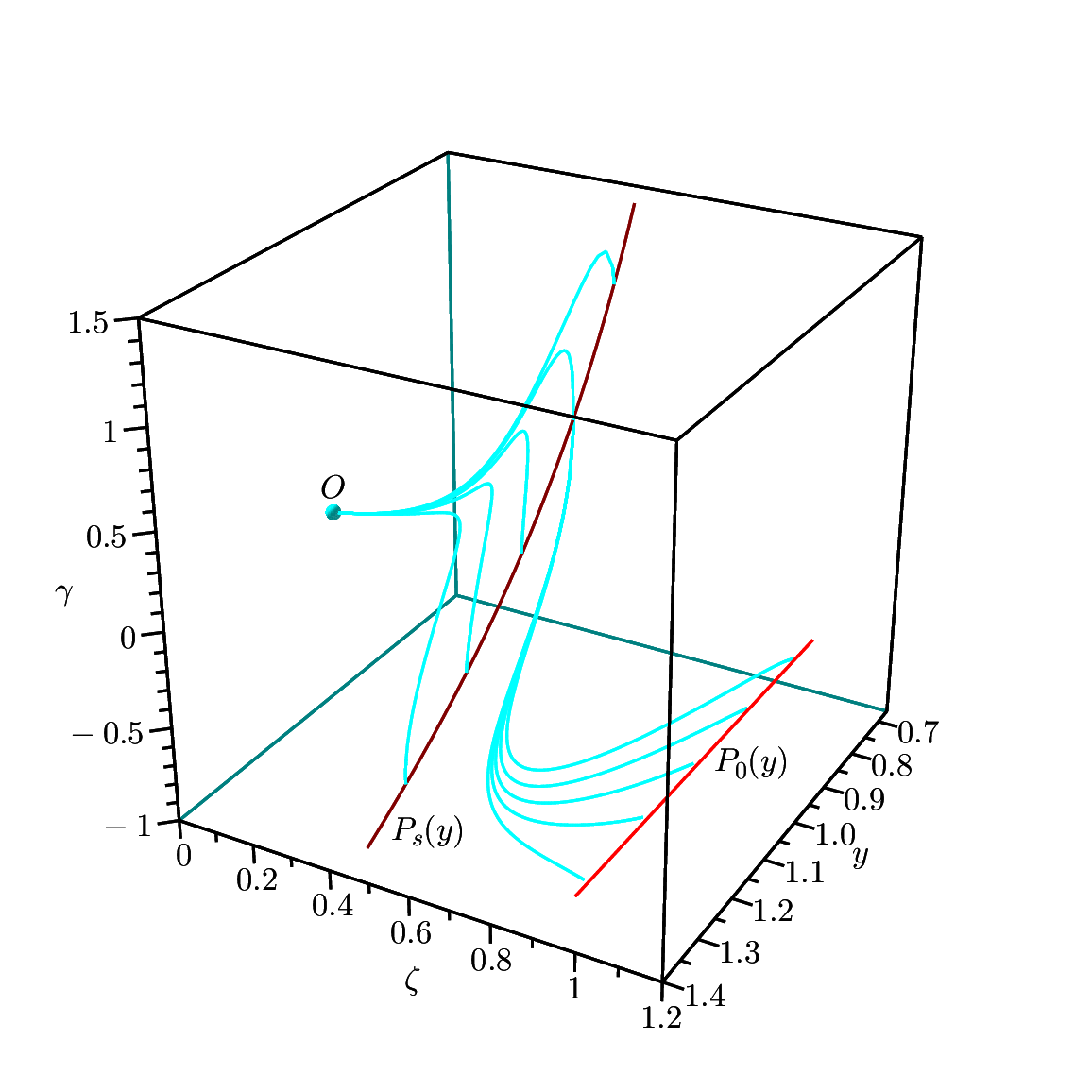}
  \caption{Numerical integration of the autonomous system (\ref{ode1}-\ref{ode3}) with $V_0=-1$ and $\kappa=1/\sqrt{2}$. Integral curves from the initial point $\mathcal{O}$ ($x=0$) first reach the singular curve $\mathcal{P}_s(y)$ ($x=x_*$, $\zeta=1/2$), and then they bifurcate and continue to move towards the line $\mathcal{P}_0(y)$ ($x=+\infty$). The plot shows bifurcation drawn from one sampling point on $\mathcal{P}_s(y)$. }
  \label{fig5}
\end{figure}

Starting from this section, we investigate the embedding of our 4-dimensional solutions into higher dimensions \cite{Cai04,Mignemi89} via the relation (\ref{metric4n}), and determine whether the basic picture of 4-dimensional naked singularity spacetimes illustrated in Fig. \ref{fig4} can carry over to higher dimensions.

For simplicity, we assume $\mathcal{K} = S^n$ or $T^n$ in the full metric (\ref{metric4n}) and fix the parameter $\kappa$ as
\begin{align}
 \kappa &= \left\{
 \begin{array}{ll}
 > 0, &\ \text{if } V_0 = 0 \phantom{-}\ \ (R_n=0,\ \mathcal{K}=T^n), \\
 \sqrt{\frac{n}{n+2}} < 1, &\ \text{if } V_0 = -1 \ \ (R_n=4,\ \mathcal{K}=S^n).
 \end{array} \right.
 \label{kappa}
\end{align}
In both situations, one can specify values of $\kappa$ that lie within the required bounds (\ref{kNS}) and (\ref{kNSV}) for naked singularities to occur in 4 dimensions.

First we consider the uplift of the solutions with $V_0=-1$. We will show that the resulting spacetimes inherit major properties of the lower-dimensional ones, and thus can represent formation of naked singularities as well (cf. Fig. \ref{fig4}). To make our calculation concrete, we focus on the simplest case with $\mathcal{K}=S^2$, $\kappa=1/\sqrt{2}$ (cf. (\ref{kappa}) with $n=2$), and take
\begin{equation} \label{dsn2}
 \rmd s^2_{n=2} = \frac{1}{2} \rmd \bar{\Omega}^2
\end{equation}
where $\rmd \bar{\Omega}^2$ denotes the metric of a unit 2-sphere. An extra factor $1/2$ above is added so that the Ricci scalar of $\mathcal{K}$ obeys $R_{n=2}=4$ (cf. (\ref{kappa})). Under the assumptions, the metric (\ref{metric4n}) becomes
\begin{equation} \label{metric6}
 \rmd \hat{s}^2_6 = \rme^{\sqrt{2}\phi} \left(-g \tilde{g}\,\rmd u^2 - 2 g\,\rmd u\,\rmd r + r^2 \rmd\Omega^2\right) + \frac{\rme^{-\sqrt{2}\phi}}{2}\, \rmd\bar\Omega^2,
\end{equation}
which acquires the spatial symmetry of $S^2\times S^2$.

\subsection{Apparent horizon}

Analogous to the metric (\ref{metric4}), the radial null geodesics of the metric (\ref{metric6}) are determined by $u=\text{const}$ for outgoing null rays and
\begin{equation}
 \frac{\rmd r}{\rmd u} = -\frac{\tilde{g}}{2},
\end{equation}
for incoming null rays (for construction of double-null coordinates, see \cite{Christo94}, Sec. 4). They correspond respectively to the future-directed null vector fields $l$ and $n$:
\begin{equation}
 l = \sqrt{\frac{\tilde{g}}{g}}\, \partial_r,
 \qquad n = \frac{2}{\sqrt{g\tilde{g}}}\, \partial_u - \sqrt{\frac{\tilde{g}}{g}}\, \partial_r,
 \qquad \rmd s^2(l,n)=-2,
\end{equation}
which are normal to a 4-dimensional spacelike surface $\Sigma=S^2\times S^2$ parameterized by $r,u=\text{const}$. The null second fundamental forms $\chi$ and $\underline\chi$ of the surface $\Sigma$ are defined by
\begin{equation}
 \hat\chi(X,Y) := \rmd\hat{s}^2_6(\hat\nabla_X l,Y),
 \qquad \hat{\underline\chi}(X,Y) := \rmd\hat{s}^2_6(\hat\nabla_X n,Y),
\end{equation}
for two arbitrary vectors $X$ and $Y$ tangent to $\Sigma$. Their traces measure the expansions of the radial null geodesic congruences along $l$ and $n$. In this setting, an apparent horizon (marginally outer trapped surface \cite{Anderson09,Galloway08,Galloway06,Chrusciel10,Senovilla11}) refers to a surface $\Sigma$ such that
\begin{equation}
 \text{tr}\hat\chi \big|_\Sigma = 0, \qquad \text{tr}\hat{\underline\chi} \big|_\Sigma \leq 0.
\end{equation}
Thus it boils down to calculating these two quantities (see Theorem \ref{thm}). The result reads
\begin{equation}
 \text{tr}\hat\chi = \hat{g}^{\alpha\beta} \chi_{\alpha\beta}
 = \frac{2}{r}\sqrt{\frac{\tilde{g}}{g}},
 \qquad \text{tr}\hat{\underline\chi} = \hat{g}^{\alpha\beta} \underline\chi_{\alpha\beta}
 = -\frac{2}{r}\sqrt{\frac{\tilde{g}}{g}},
\end{equation}
for general metric functions $g=g(u,r)$, $\tilde{g}=\tilde{g}(u,r)$, and $\phi=\phi(u,r)$. Remarkably, the same expressions also hold for the metric $\rmd s^2$ ($\Sigma=S^2$) itself, as well as when $\mathcal{K}=T^1$ ($\Sigma=S^2\times T^1$, see (\ref{metric5})). This is due to certain key cancellations involving $\phi$ that is enabled by the Kaluza-Klein reduction.  Hence identically to the situation in 4 dimensions (cf. (\ref{mass})), absence of the apparent horizon is signaled by $\tilde{g}/g=y\neq 0$ in the lifted spacetime. For similar discussions on invariance of trapping in the Kaluza-Klein dimensional reduction, one can see \cite{Senovilla02,Paetz13}.

\begin{thm} \label{thm}
The apparent horizon of the 5+1 dimensional metric $\rmd \hat{s}^2_6$ is a natural lift of the apparent horizon of the constituent 3+1 dimensional metric $\rmd s^2$.
\end{thm}

\begin{proof}

We introduce spherical coordinates to $\rmd\hat{s}^2_6=\hat{g}_{\hat\mu\hat\nu} \rmd x^{\hat\mu} \rmd x^{\hat\nu}$ such that $\rmd \Omega^2 = \rmd\theta^2+\sin^2\theta\rmd\varphi^2$ and $\rmd\bar\Omega^2 = \rmd\bar\theta^2+\sin^2\bar\theta\rmd\bar\varphi^2$. The relevant Christoffel symbols are
\begin{equation}
 \hat\Gamma^\theta_{\theta r} = \frac{1}{r} + \frac{\sqrt{2}}{4}\phi_{,r} = \hat\Gamma^\varphi_{\varphi r}, \qquad
 \hat\Gamma^{\bar\theta}_{\bar\theta r} = -\frac{\sqrt{2}}{4}\phi_{,r} = \hat\Gamma^{\bar\varphi}_{\bar\varphi r},
\end{equation}
with all non-diagonal $\hat\Gamma^\alpha_{\beta r}=0$ and $\alpha,\beta\in\{\theta,\varphi,\bar\theta,\bar\varphi\}$. Then we have
\begin{align}
 \hat\nabla_\theta l &= \sqrt{\frac{\tilde g}{g}}\,\hat\Gamma^\theta_{\theta r} \partial_\theta
 = \sqrt{\frac{\tilde g}{g}} \left(\frac{1}{r} + \frac{\sqrt{2}}{4}\phi_{,r} \right) \!\partial_\theta, \\
 \hat\nabla_\varphi l &= \sqrt{\frac{\tilde g}{g}}\,\hat\Gamma^\varphi_{\varphi r} \partial_\varphi
 = \sqrt{\frac{\tilde g}{g}} \left(\frac{1}{r} + \frac{\sqrt{2}}{4}\phi_{,r} \right) \!\partial_\varphi, \\
 \hat\nabla_{\bar\theta} l &= \sqrt{\frac{\tilde g}{g}}\,\hat\Gamma^{\bar\theta}_{\bar\theta r} \partial_{\bar\theta}
 = -\sqrt{\frac{\tilde g}{g}} \frac{\sqrt{2}}{4}\phi_{,r} \partial_{\bar\theta}, \\
 \hat\nabla_{\bar\varphi} l &= \sqrt{\frac{\tilde g}{g}}\,\hat\Gamma^{\bar\varphi}_{\bar\varphi r} \partial_{\bar\varphi}
 = -\sqrt{\frac{\tilde g}{g}} \frac{\sqrt{2}}{4}\phi_{,r} \partial_{\bar\varphi}.
\end{align}
Thereby, we calculate the trace as
\begin{align}
 \text{tr}\hat\chi &= \hat{g}^{\theta\theta} \rmd\hat{s}^2_6(\hat\nabla_\theta l,\partial_\theta)
 + \hat{g}^{\varphi\varphi} \rmd\hat{s}^2_6(\hat\nabla_\varphi l,\partial_\varphi)
 + \hat{g}^{\bar\theta\bar\theta} \rmd\hat{s}^2_6(\hat\nabla_{\bar\theta} l,\partial_{\bar\theta})
 + \hat{g}^{\bar\varphi\bar\varphi} \rmd\hat{s}^2_6(\hat\nabla_{\bar\varphi} l,\partial_{\bar\varphi}) \nonumber \\
 &= 2\sqrt{\frac{\tilde g}{g}} \left(\frac{1}{r} + \frac{\sqrt{2}}{4}\phi_{,r} \right)
 + 2\sqrt{\frac{\tilde g}{g}} \left(-\frac{\sqrt{2}}{4}\phi_{,r}\right) \nonumber \\
 &= \frac{2}{r} \sqrt{\frac{\tilde g}{g}} \,.
\end{align}
Similarly, from
\begin{equation}
 \hat\Gamma^\theta_{\theta u} = \frac{\sqrt{2}}{4}\phi_{,u} = \hat\Gamma^\varphi_{\varphi r},
 \qquad \hat\Gamma^{\bar\theta}_{\bar\theta u} = -\frac{\sqrt{2}}{4}\phi_{,u} = \hat\Gamma^{\bar\varphi}_{\bar\varphi u},
\end{equation}
with all non-diagonal $\hat\Gamma^\alpha_{\beta u}=0$, we obtain $\text{tr}\hat{\underline\chi} = -2\sqrt{\tilde g/g}/r$. Through the same procedure, one can verify the same expressions for $\rmd s^2 = g_{\mu\nu} \rmd x^\mu \rmd x^\nu$:
\begin{align}
 \text{tr}\chi &= g^{\theta\theta} \rmd s^2(\nabla_\theta l,\partial_\theta)
 + g^{\varphi\varphi} \rmd s^2 (\nabla_\varphi l,\partial_\varphi) = \frac{2}{r} \sqrt{\frac{\tilde g}{g}} \,, \\
 \text{tr}\underline\chi &= g^{\theta\theta} \rmd s^2(\nabla_\theta n,\partial_\theta)
 + g^{\varphi\varphi} \rmd s^2 (\nabla_\varphi n,\partial_\varphi) = -\frac{2}{r} \sqrt{\frac{\tilde g}{g}} \,,
\end{align}
by using the fact that $\nabla_\theta \partial_r = r^{-1}\partial_\theta$, $\nabla_\varphi \partial_r = r^{-1}\partial_\varphi$, and $\nabla_\theta \partial_u = 0 = \nabla_\varphi \partial_u$.

\end{proof}

\subsection{Kretschmann scalar}

Much like its 4-dimensional counterpart, the lifted spacetime can remain nonsingular for $u<0$, both at the center and beyond. A scalar-valued curvature singularity first occurs at $r=0$, $u=0$. To see this, one should examine the Kretschmann scalar $\hat{K}=\hat{R}^{\mu\nu\rho\sigma}\hat{R}_{\mu\nu\rho\sigma}$ particularly at three key locations in the spacetime, i.e., $x=0$, $+\infty$ and $x_*$. In terms of $y$, $\zeta$ and $\gamma$, the full expression of $\hat{K}$ is given by
\begin{align}
 \hat{K} &= 12 \big\{\big[ -6(2\zeta-1)^2\gamma^4 - 2\sqrt{2}(9\zeta-4)(2\zeta-1)\gamma^3
 -\zeta(25\zeta-12)\gamma^2 \nonumber \\
 & -4\sqrt{2}\zeta^2\gamma - 9\zeta^2 + 16\zeta - 8 \big] y^2 + \big[-8(2\zeta-1)(\zeta-1)\gamma^2 \nonumber \\
 & -4\sqrt{2}\zeta(\zeta-1)\gamma + 16(\zeta-1)^2 \big]y - 8(\zeta-1)^2 \big\} \nonumber \\
 & \big/ r^2(\zeta-1)\big[2\gamma(\gamma+\sqrt{2})(2\zeta-1)y - (\zeta-2)y+2(\zeta-1)\big], \label{K6}
\end{align}
where we have used the equations (\ref{ode1}-\ref{ode3}) to eliminate the derivatives of $y$, $\zeta$ and $\gamma$.

Near the initial point $\mathcal{O}$ with $x=0$, the system (\ref{ode1}-\ref{ode3}) admits the Taylor series solution (\ref{seriesT}) with a parameter $c_T<\kappa^2/3$ ($c_T=\kappa^2/3$ for $V_0=0$). Plugging it into (\ref{K6}), we obtain
\begin{equation}
 \hat{K} = \frac{18(1-4c_T)^2}{(1-6c_T)u^2} \left(1 + 2x + O(x^2)\right),
 \qquad c_T<\frac{\kappa^2}{3}=\frac{1}{6},
\end{equation}
which stays finite and continuous at $r=0$ for finite $u<0$. Similarly, near an ending point $\mathcal{P}_0(y_0)$ with $x=+\infty$, the solution satisfies the asymptotic limits (\ref{yzgP0}), which results in
\begin{equation} \label{KP0}
 \hat{K} \rightarrow \frac{12y_0^2}{(3y_0-2)r^2}, \ \ \text{as}\ \ s\rightarrow +\infty\ (u\rightarrow 0-).
\end{equation}
Hence, the scalar $\hat{K}$ diverges at $r=0$, but remains bounded elsewhere along $u=0$ (the Cauchy horizon). Related to this, the same fall-off behavior also holds for other fixed $u<0$:
\begin{equation} \label{Ku}
 \hat{K} = \frac{12y_0^2}{(3y_0-2)r^2} + o(r^{-2}) \rightarrow 0,\ \ \textrm{as}\ \ r\rightarrow +\infty.
\end{equation}

Now move on to $x=x_*$. The expression of (\ref{K6}) is continuous at the singular point $\mathcal{P}_s(y_*)$:
\begin{equation}
 y = y_*, \ \zeta = \frac{1}{2},
 \ \gamma = \frac{1-y_*}{y_*\kappa^3} = \frac{2\sqrt{2}(1-y_*)}{y_*},
 \ y_* > \frac{1}{1+\kappa^2} = \frac{2}{3}.
\end{equation}
Hence we obtain
\begin{equation} \label{KPs}
 \hat{K} \rightarrow \frac{12y_*^2}{(3y_*-2)r^2}, \ \ \text{as}\ \ s\rightarrow s_*,
\end{equation}
which resembles the limit (\ref{KP0}).

For other fixed values of $x>0$, the equation (\ref{K6}) implies that
\begin{equation} \label{Kx}
 \hat{K} \propto \frac{1}{r^2} \rightarrow 0, \ \ \textrm{as}\ \ r\rightarrow +\infty.
\end{equation}
The boundedness of $\hat{K}$ for $r\neq 0$ ensues from the local smoothness of the solution curves, and is also verifiable by numerical calculation.

\subsection{Asymptotic local flatness}

With $\kappa=1/\sqrt{2}$ (cf. (\ref{kappa}) with $n=2$), the limits in (\ref{yzgP0}) and (\ref{phi}) imply that
\begin{align}
 \frac{g\tilde{g}}{x} \rightarrow \frac{4c_2^2}{y_0},
 \ \frac{g}{\sqrt{x}} \rightarrow -\frac{2c_2}{y_0},
 \ x \rme^{\sqrt{2}\phi} \rightarrow \frac{\sqrt{3y_0-2}}{-2u},
 \ \ \text{as}\ \ x\rightarrow +\infty.
\end{align}
Introducing new coordinates
\begin{equation} \label{rttilde}
 \tilde{r}^2 = r, \qquad \tilde{t} - \tilde{r} = \tilde{u} = 2c_2 \sqrt{-u}<0,
\end{equation}
thus for finite $u<0$ and $r$ (equivalently $x$) being large enough, we have
\begin{align}
 &\rmd \hat{s}^2 = \rme^{\sqrt{2}\phi}\left(-g(x) \tilde{g}(x)\,\rmd u^2 - 2 g(x)\,\rmd u\,\rmd r + r^2 \rmd\Omega^2\right) + \frac{\rme^{-\sqrt{2}\phi}}{2}\, \rmd\bar\Omega^2
 \\
 &\approx -\frac{2\sqrt{3y_0-2}}{y_0}
 \left(\frac{c_2^2\rmd u^2}{-u} - \frac{2c_2\rmd u\rmd\tilde{r}}{\sqrt{-u}}\right)
 + \frac{\sqrt{3y_0-2}}{2}\, \tilde{r}^2 \rmd\Omega^2
 + \frac{\tilde{r}^2}{\sqrt{3y_0-2}}\, \rmd\bar\Omega^2 \label{approx}
 \\
 &= -\frac{2\sqrt{3y_0-2}}{y_0}
 \left(\rmd\tilde{u}^2 + 2\rmd\tilde{u}\rmd\tilde{r} \right)
 + \tilde{r}^2 \left(\frac{\sqrt{3y_0-2}}{2}\, \rmd\Omega^2
 + \frac{1}{\sqrt{3y_0-2}}\, \rmd\bar\Omega^2 \right)
 \\
 &= \frac{2\sqrt{3y_0-2}}{y_0}
 \left(-\rmd\tilde{t}^2 + \rmd\tilde{r}^2 \right)
 + \tilde{r}^2 \left(\frac{\sqrt{3y_0-2}}{2}\, \rmd\Omega^2
 + \frac{1}{\sqrt{3y_0-2}}\, \rmd\bar\Omega^2 \right), \label{approxMin}
\end{align}
where we only keep dominant terms in (\ref{approx}). Then note that by stereographic projection, $\tilde{r}^2\rmd\Omega^2 = (\rmd \tilde{x}^2+\rmd \tilde{y}^2)/[1+(\tilde{x}^2+\tilde{y}^2)/4\tilde{r}^2]^2 \rightarrow \rmd \tilde{x}^2+\rmd \tilde{y}^2$ for local $\tilde{x}$ and $\tilde{y}$ as $\tilde{r}\rightarrow +\infty$. Combining these asymptotic expressions with the decaying Kretschmann scalar in (\ref{Ku}), we comment that the lifted spacetime, by suitable truncation \cite{Christo94}, is asymptotically locally flat with non-trivial topology $S^2\times S^2$ at spatial infinity. This significantly improves from the non-trivial asymptotic behavior of the 4-dimensional spacetime discussed in Sec. \ref{sec:4D}.

\subsection{Strength of singularities}

In the asymptotic metric (\ref{approxMin}) as $x\rightarrow +\infty$, the coordinate $\tilde r$, modulo a factor, essentially plays the role of the ``area'' radius. Together with (\ref{KP0}) and (\ref{rttilde}), we have
\begin{equation} \label{Kr4}
\hat{K}\propto \frac{1}{r^2} \propto \frac{1}{\tilde{r}^4}.
\end{equation}
The power $4$ here is a crucial index. We use it to measure the strength of singularities.

Now recall the 5+1 dimensional Schwarzschild metric
\begin{equation}
\rmd \hat{s}^2_{Sch} = -\left(1-\frac{2m}{\tilde{r}^3}\right) \rmd t^2 + \left( 1-\frac{2m}{\tilde{r}^3} \right)^{-1} \rmd \tilde{r}^2 + \tilde{r}^2 \rmd \Omega_4^2,
\end{equation}
with the Kretschmann scalar given by
\begin{equation}
 \hat K_{Sch} = \frac{960 m^2}{\tilde{r}^{10}}.
\end{equation}
By comparing the power index of $\hat K$ near the center $\tilde{r}=0$, it can be said that the central naked singularity in our solution is less singular than the one in the Schwarzschild spacetime (for the 3+1 dimensional Schwarzschild metric, $K_{Sch} = 48m^2/r^6$).

The $1/\tilde{r}^4$ blow-up rate of $\hat K$ can also show up naturally in metrics that contain conical singularities. One quick example happens to be the metric (\ref{approxMin}) extended inward to $\tilde{r}=0$. This represents a borderline scenario. Therefore, one can think of our naked singularity solution as a threshold between regular and strongly singular (black hole) solutions to the vacuum Einstein equations.

\subsection{Homothetic Killing vector}

With the assumption of (\ref{dsn2}) and $\kappa=1/\sqrt{2}$, one can check that the lifted metric $\rmd\hat{s}^2_6=\hat{g}_{\hat\mu\hat\nu} \rmd x^{\hat\mu} \rmd x^{\hat\nu}$ possesses the same homothetic Killing vector $\xi$ as $\rmd s^2=g_{\mu\nu} \rmd x^\mu \rmd x^\nu$ \cite{Christo94,Brady95}:
\begin{equation}
 \xi = u\partial_u + r\partial_r, \qquad L_\xi g_{\mu\nu} = 2 g_{\mu\nu},
 \qquad L_\xi \hat{g}_{\hat\mu\hat\nu} = \hat{g}_{\hat\mu\hat\nu}.
\end{equation}
Therefore, the 6-dimensional naked singularity spacetimes with spherical extra dimensions are continuously self-similar.

\section{Comment on 5 dimensions} \label{sec:5D}

From the previous section, one would anticipate that lifting the solutions with $V_0=0$ for the massless scalar field might also generate new naked singularity solutions in higher dimensions (see \cite{Feinstein06} for discussions on lifting solutions with $\kappa=0$). However, we will show that this turns out to be false. The reason lies in that, unlike the 6-dimensional example, the Kretschmann scalar blows up at $x=x_*$, i.e., along the past light cone of the central singularity. Hence in the resulting spacetime, a curvature singularity can pre-exist before the central singularity emerges later at $u=0$.

Again for concreteness of calculation, we focus on the simplest case $\mathcal{K}=T^1(=S^1)$ in 5 dimensions and take
\begin{equation} \label{dsn5}
 \rmd s^2_{n=1}=(\rmd x^1)^2
\end{equation}
with $x^1$ the local coordinate on $\mathcal{K}$. The metric reads
\begin{equation} \label{metric5}
  \rmd \hat{s}^2_5 = \rme^{\frac{2}{\sqrt{3}}\phi}\left(-g \tilde{g}\,\rmd u^2 - 2 g\,\rmd u\,\rmd r + r^2 \rmd\Omega^2\right) + \rme^{-\frac{4}{\sqrt{3}}\phi}\, (\rmd x^1)^2.
\end{equation}
Using (\ref{eqaml}), we can remove the unknown $y$ in the equation (\ref{ode3}) and obtain a reduced system for $\zeta$ and $\gamma$. It can also be verified that the equations (\ref{ode2}-\ref{eqaml}) together imply (\ref{ode1}). To solve this 2-dimensional autonomous system in a neighborhood of the singular point $\mathcal{P}_s(\frac{1}{1+\kappa^2})$, one can apply the same linearization procedure as in (\ref{ts}-\ref{eigen_Ps}) and obtain (\cite{Christo94}, Sec. 2; for a comparison of notations, see the Appendix \ref{app:Christo})
\begin{align}
 \zeta &= \frac{1}{2} + a_2 \rme^{(1-\kappa^2)t} + \cdots, \\
 \gamma &= \frac{1}{\kappa} + a_1\rme^{\kappa^2 t}
 + \frac{4a_2(1+\kappa^2)}{\kappa^3(1-2\kappa^2)}\, \rme^{(1-\kappa^2)t} + \cdots,
\end{align}
where the variable $t$ is defined by (\ref{ts}) and $a_{1,2}$ are two nonzero parameters. Owing to these asymptotic solutions for $t\rightarrow -\infty$ ($s\rightarrow s_*$), the Kretschmann scalar (written in terms of $\zeta$ and $\gamma$ without derivatives)
\begin{align}
\hat{K} &= \big[-16( 4\sqrt{3}\kappa-3{\kappa}^{2}-7)(2\zeta-1)^{3}{\gamma}^{4}
-32(4\sqrt{3}\zeta{\kappa}^{2}-3\zeta{\kappa}^{3} -6\sqrt{3}\zeta-7\zeta\kappa
\nonumber \\
& +4\sqrt{3}) (2\zeta-1) ^{2}{\gamma}^{3}-4(20\sqrt{3}{\zeta}^{2}{\kappa}^{3}-15{\zeta}^{2}{
\kappa}^{4}-84\sqrt{3}{\zeta}^{2}\kappa-109{\zeta}^{2}{\kappa}^{2}
\nonumber \\
& +48\sqrt{3}\zeta\kappa+108 \zeta {\kappa}^{2}-78{\zeta}^{2}-36{\kappa}^{2}+108\zeta-36) (2\zeta-1){\gamma}^{2} - 4(4\sqrt{3}{\zeta}^{2}{\kappa}^{3}
\nonumber \\
& -3{\zeta}^{2}{\kappa}^{4}-48\sqrt{3}{\zeta}^{2}\kappa-99{\zeta}^{2}{\kappa}^{2}+24\sqrt{3}\zeta\kappa+120\zeta{\kappa}^{2}-96{\zeta}^{2}-36{\kappa}^{2}+120\zeta
\nonumber \\
& -36 )\zeta\kappa\gamma+4( 4\sqrt{3}\zeta\kappa+15\zeta{\kappa}^{2}-9{\kappa}^{2}+15\zeta-9 ) {\zeta}^{2}{\kappa}^{2} \big] / 3 r^{4-4\kappa/\sqrt{3}} x^{4\kappa/\sqrt{3}}
\nonumber \\
&  \times \rme^{4\bar{h}/\sqrt{3}} (2\zeta-1)[(2\zeta-1)(\gamma+2\kappa)\gamma+(\kappa^2-1)\zeta+1]^2
\end{align}
diverges as follows:
\begin{equation} \label{K5}
 \lim_{t\rightarrow -\infty} \rme^{(1-2\kappa^2)t} \hat{K} \propto \frac{1}{r^{4-4\kappa/\sqrt{3}}},
 \qquad 0<\kappa<\frac{1}{\sqrt{3}}.
\end{equation}
One can also confirm the divergence of $\hat{K}$ at $x=x_*$ via numerical integrations along individual solution curves.

As an aside, we note that the lifted metric $\rmd \hat{s}^2_5$ can no longer admit the homothetic Killing vector $\xi=u\partial_u+r\partial_r$ unless $\kappa=1/\sqrt{3}$, of which the value falls out of the required bound $0<\kappa^2<1/3$ for emergence of naked singularities in 4 dimensions.

At first glance, the inconsistency of spacetime regularity (cf. (\ref{K5})) in the Kaluza-Klein dimensional reduction may appear puzzling. One physical way to resolve it is by changing the conformal frame for $\rmd s^2$ \cite{Faraoni99}, which means that one should instead look at the 4-dimensional projection of the full metric $\rmd \hat{s}^2_5$, i.e., $\exp(2\phi/\sqrt{3})\rmd s^2$. For this re-scaled metric with the dilaton $\phi$ involved, direct calculation shows that the Kretschmann scalar diverges at $x=x_*$ ($K$ continuous for $\rmd s^2$ except at $r=0$, $u=0$; $K\propto 1/r^4$) in accordance with its 5-dimensional counterpart.

\section{Concluding Remarks} \label{sec:conclu}

In the context of the Kaluza-Klein dimensional reduction, we have constructed a 2-parameter family of 6-dimensional naked singularity solutions by lifting the 4-dimensional naked singularity solutions of the Einstein-scalar field system with a negative exponential potential. Despite its success, the same construction does not come through in the case of a vanishing potential due to a diverging Kretschmann invariant along the incoming null ray $x=x_*$ (hence in the initial data), which raises a caveat for utilizing this type of embedding. Regarding the 6-dimensional solutions, it may draw some concern that the spatial infinity has a non-trivial topology $S^2\times S^2$. However, we should emphasize that the local formation of a singularity near the center does not depend on the spacetime region far from it. Hence it would be interesting to investigate whether the central region of the spacetime can be truncated and matched to an appropriate surrounding region that is, for instance, asymptotically Minkowski ($S^4$) or asymptotically Kaluza-Klein (compact extra dimensions). Related to this, we recall that truncation is necessary for Christodoulou's solutions \cite{Christo94}, in order to obtain asymptotically flat initial data. As a final remark, it should be noted that though we have restricted the discussion to 6 and 5 dimensions, our treatment can be generalized to other values of $4+n\geq 6$.

\appendix

\section{The vacuum Einstein equations in 6 dimensions} \label{app:EVE6}

We write down explicitly the vacuum Einstein equations $\hat R_{\hat\mu\hat\nu}=0$ for the 6-dimensional metric (\ref{metric6}) with $\rmd \Omega^2 = \rmd\theta^2+\sin^2\theta\rmd\varphi^2$ and $\rmd\bar\Omega^2 = \rmd\bar\theta^2+\sin^2\bar\theta\rmd\bar\varphi^2$. The non-vanishing components of the Ricci tensor are
\begin{align}
 \hat R_{rr} &= 2(\ln g)_{,r}/r - 2(\phi_{,r})^2, \\
 \hat R_{uu} - \tilde g \hat R_{ur} &= g(\tilde{g}/g)_{,u}/r - 2[(\phi_{,u})^2 - \tilde{g} \phi_{,u}\phi_{,r}], \\
 \hat R_{\theta\theta} &= -\frac{r}{\sqrt{2}g}\left[r^{-1} (r^2 \tilde{g}\phi_{,r})_{,r} - 2\phi_{,u} -2r\phi_{,ru}\right] - \frac{(r\tilde{g})_{,r}}{g} + 1, \\
 \hat R_{\bar\theta\bar\theta} &= \frac{\rme^{-2\sqrt{2}\phi}}{2\sqrt{2}rg}
 \left[ r^{-1} (r^2 \tilde{g}\phi_{,r})_{,r} - 2\phi_{,u} - 2r\phi_{,ru} + 2\sqrt{2} r g \rme^{2\sqrt{2}\phi} \right], \\
 \hat R_{\varphi\varphi} &= \sin^2\theta \hat R_{\theta\theta}, \qquad
 \hat R_{\bar\varphi\bar\varphi} = \sin^2\bar\theta \hat R_{\bar\theta\bar\theta}, \\
 \hat R_{ur} &= \frac{\tilde{g}\phi_{,rr}}{2\sqrt{2}} - \frac{\phi_{,ur}}{\sqrt{2}} - \frac{\phi_{,u}\phi_{,r}}{2} + \frac{\tilde{g}\phi_{,r}}{\sqrt{2}r} + \frac{\tilde{g}_{,r} \phi_{,r}}{2\sqrt{2}} - \frac{\phi_{,u}}{\sqrt{2}r} + \frac{g_{,rr} \tilde{g}}{2g}
 \nonumber \\
 & + \frac{\tilde{g}_{,rr}}{2} - \frac{g_{,ur}}{g} + \frac{g_{,r}\tilde{g}_{,r}}{2g} - \frac{\tilde{g}(g_{,r})^2}{2g^2} + \frac{g_{,r}g_{,u}}{g^2} + \frac{g_{,r} \tilde{g}}{gr} + \frac{\tilde{g}_{,r}}{r}.
\end{align}
Given $\kappa=1/\sqrt{2}$ (cf. (\ref{kappa}) with $n=2$), it is straightforward to check that the equations of motion (\ref{feq1}-\ref{feq4}) for the Einsteins-scalar field system (\ref{Lag}-\ref{Vphi}) imply $\hat R_{\hat\mu\hat\nu}=0$ above. Hence all solutions of the lower-dimensional reduced system automatically produce solutions of the higher-dimensional one.

\section{Christodoulou's notation} \label{app:Christo}

One can translate Brady's notation that we have been using to Christodoulou's \cite{Christo94} through
\begin{equation}
 y^{-1}=\rme^{2\lambda}, \qquad z^{-1}=2(1-\beta), \qquad \gamma = \theta, \qquad \kappa = k.
\end{equation}
In terms of the latter, the system (\ref{ode1}-\ref{ode3}) reads
\begin{align}
 \frac{\rmd \lambda}{\rmd s} =& -\frac{(\theta+k)(2\beta\theta-\theta+k)}{2(1-\beta)}, \label{odec1}\\
 \frac{\rmd \beta}{\rmd s} =& 1-k^2 - \left[(\theta+k)^2+1-k^2\right]\beta,  \label{odec2}\\
 \frac{\rmd \theta}{\rmd s} =& \frac{k}{\beta}(k\theta-1)+\left[(\theta+k)^2-(1+k^2)\right]\theta
 \nonumber \\
 &+ \frac{1-\beta}{k\beta}
 \left[ 1+k^2 + \frac{\beta}{1-\beta}(\theta+k)^2 - \rme^{2\lambda}\right]. \label{odec3}
\end{align}
Also the constraint (\ref{eqaml}) for the massless scalar field becomes
\begin{equation} \label{eqamlc}
 \rme^{2\lambda} = 1+k^2 + \frac{\beta}{1-\beta}(\theta+k)^2.
\end{equation}
Using this equation, we can remove $\lambda$ in (\ref{odec3}) and obtain a reduced system
\begin{align}
 \frac{\rmd \beta}{\rmd s} &= 1-k^2 - \left[(\theta+k)^2+1-k^2\right]\beta, \\
 \frac{\rmd \theta}{\rmd s} &= \frac{k}{\beta}(k\theta-1)+\left[(\theta+k)^2-(1+k^2)\right] \theta,
\end{align}
which recovers the equations (0.27a,b) in \cite{Christo94}. It is straightforward to verify that if the above system holds for $\beta$ and $\theta$, the function $\lambda$, as determined by (\ref{eqamlc}), automatically satisfies (\ref{odec1}).


\subsection*{Acknowledgment}
We are especially grateful to Hong L\"u who provided us with a personal note and his paper \cite{Bremer99} on the Kaluza-Klein reduction and suggested the embedding problem into higher dimensions. We extend our gratitude to Daniel Finley, Sijie Gao, Christopher Pope, and Bohua Zhan for helpful discussions. We also thank Luis Lehner and the anonymous referee for valuable suggestions and comments on earlier versions of the manuscript. The project was supported by China Postdoctoral Science Foundation, NSFC grants 11475024, 11235003, and 91636111. XA also acknowledges AIP Fund from Rutgers University.


\begin{thebibliography}{100}


\bibitem{Abrahams93}
  A.M. Abrahams and C.R. Evans
  {\it Critical behavior and scaling in vacuum axisymmetric gravitational collapse},
  Phys. Rev. Lett. {\bf 70}, 2980 (1993).

\bibitem{An12}
  X. An,
  {\it Formation of trapped surfaces from past null infinity},
  arXiv:1207.5271 (2012).

\bibitem{An17}
  X. An and J. Luk,
  {\it Trapped surfaces in vacuum arising dynamically from mild incoming radiation},
  Adv. Theor. Math. Phys. {\bf 21}, 1 (2017). 

\bibitem{Anderson09}
  L. Anderson and J. Metzger,
  {\it The area of horizons and the trapped region},
  Comm. Math. Phys. {\bf 290}, 941 (2009).

\bibitem{Bedjaoui10}
  N. Bedjaoui, P.G. LeFloch, J.M. Mart\'{\i}n-Garc\'{\i}a, and J. Novak,
  {\it Existence of naked singularities in the Brans--Dicke theory of gravitation. An analytical and numerical study},
  Class. Quantum Grav. {\bf 27}, 245010 (2010).

\bibitem{Bizon05}
  P. Bizo\'{n}, T. Chmaj, and B.G. Schmidt,
  {\it Critical behavior in vacuum gravitational collapse in 4+1 dimensions},
  Phys. Rev. Lett.  {\bf 95}, 071102 (2005).

\bibitem{Bizon11}
  P. Bizo\'{n} and A. Rostworowski, 
  {\it Weakly Turbulent Instability of Anti-de Sitter Spacetime},
  Phys. Rev. Lett.  {\bf 107}, 031102 (2011).

\bibitem{Bizon00}
  P. Bizo\'{n} and A. Wasserman,
  {\it Self-similar spherically symmetric wave maps coupled to gravity},
  Phys. Rev. D {\bf 62}, 084031 (2000).

\bibitem{Brady95}
  P.R. Brady,
  {\it Self-similar scalar field collapse: Naked singularities and critical behavior},
  Phys. Rev. D {\bf 51}, 4168 (1995).

\bibitem{Brady97}
  P.R. Brady, C.M. Chambers, and S.M.C.V. Gon\c{c}alves,
  {\it Phases of massive scalar field collapse},
  Phys. Rev. D {\bf 56}, R6057 (1997).

\bibitem{Bremer99}
  M.S. Bremer, M.J. Duff, H. L\"u, C.N. Pope, and K.S. Stelle,
  {\it Instanton cosmology and domain walls from M-theory and string theory},
  Nucl. Phys. B {\bf 543}, 321 (1999).

\bibitem{Cai04}
  R.-G. Cai and A. Wang,
  {\it Nonasymptotically AdS/dS solutions and their higher dimensional origins},
  Phys. Rev. D {\bf 70}, 084042 (2004).

\bibitem{Carr99}
  B.J. Carr and A.A. Coley,
  {\it Self-similarity in general relativity},
  Class. Quantum Grav. {\bf 16}, R31 (1999).

\bibitem{Carr05}
  B.J. Carr and A.A. Coley,
  {\it The similarity hypothesis in general relativity},
  Gen. Rel. Grav. {\bf 37}, 2165 (2005).

\bibitem{Chan95}
  K.C.K. Chan, J.H. Horne, and R.B. Mann,
  {\it Charged dilaton black holes with unusual asymptotics},
  Nucl. Phys. B {\bf 447}, 441 (1995).

\bibitem{Choptuik93}
  M.W. Choptuik,
  {\it Universality and scaling in gravitational collapse of a massless scalar field},
  Phys. Rev. Lett. {\bf 70}, 9 (1993).

\bibitem{Christo94}
  D. Christodoulou,
  {\it Examples of naked singularity formation in the gravitational collapse of a scalar field},
  Ann. Math. {\bf 140}, 607 (1994).

\bibitem{Christo99}
  D. Christodoulou,
  {\it The instability of naked singularities in the gravitational collapse of a scalar field},
  Ann. Math. {\bf 149}, 183 (1999).

\bibitem{Christo09}
  D. Christodoulou,
  {\it The Formation of Black Holes in General Relativity},
  Monographs in Mathematics, European Mathemcatical Society, Z\"urich (2009).

\bibitem{Chrusciel10}
  P.T. Chru\'{s}ciel, G.J. Galloway, and D. Pollack,
  {\it Mathematical general relativity: A sampler},
  Bull. Am. Math. Soc. {\bf 47}, 567 (2010).

\bibitem{Coley03}
  A.A. Coley,
  {\it Dynamical Systems and Cosmology},
  Kluwer Academic Publishers, Dordrecht (2003).

\bibitem{Conte08}
  R. Conte and M. Musette,
  {\it The Painlev\'{e} Handbook},
  Springer, Dordrecht (2008).

\bibitem{Dafermos05}
  M. Dafermos,
  {\it On naked singularities and the collapse of self-gravitating Higgs fields},
  Adv. Theor. Math. Phys. {\bf 9}, 575 (2005).

\bibitem{Dafermos09}
  M. Dafermos,
  {\it The evolution problem in general relativity},
  Current Developments in Mathematics, 2008, 1--66, Somerville, MA (2009).

\bibitem{Dafermos13}
  M. Dafermos,
  {\it The formation of black holes in general relativity [after D. Christodoulou]},
  Astrisque 352 (2013).

\bibitem{Dafermos08}
  M. Dafermos and I. Rodnianski,
  {\it Lectures on black holes and linear waves},
  Clay Mathematics Proceedings, Vol. 17, Amer. Math. Soc., Providence, RI (2013), pp. 97-205.

\bibitem{Duff86}
  M.J. Duff, B.E.W. Nilsson, and C.N. Pope,
  {\it Kaluza-Klein  supergravity},
  Phys. Rep. {\bf 130}, 1 (1986).

\bibitem{Emparan08}
  R. Emparan and H.S. Reall,
  {\it Black Holes in Higher Dimensions},
  Living Rev. Rel. {\bf 11}, 6 (2008).

\bibitem{Feinstein06}
  A. Feinstein,
  {\it Formation of a Black String in a Higher Dimensional Vacuum Gravitational Collapse},
  Phys. Lett. A {\bf 372}, 4337 (2008).

\bibitem{Faraoni99}
  V. Faraoni, E. Gunzig, and P. Nardone,
  {\it Conformal transformations in classical gravitational theories and in cosmology},
  Fund. Cosmic. Phys. {\bf 20}, 121 (1999).

\bibitem{Figueras16}
  P. Figueras, M. Kunesch, and S. Tunyasuvunakool,
  {\it End Point of Black Ring Instabilities and the Weak Cosmic Censorship Conjecture},
  Phys. Rev. Lett. {\bf 116}, 071102 (2016). 

\bibitem{Figueras17}
  P. Figueras, M. Kunesch, L. Lehner and S. Tunyasuvunakool,
  {\it End Point of the Ultraspinning Instability and Violation of Cosmic Censorship},
  Phys. Rev. Lett. {\bf 118}, 151103 (2017). 

\bibitem{Frolov04}
  A.V. Frolov,
  {\it Is it really naked? On cosmic censorship in string theory},
  Phys. Rev. D {\bf 70}, 104023 (2004).

\bibitem{Galloway08}
  G.J. Galloway,
  {\it Rigidity of marginally trapped surfaces and the topology of black holes},
  Comm. Anal. Geom. {\bf 16}, 217 (2008).

\bibitem{Galloway06}
  G.J. Galloway and R. Schoen,
  {\it A generalization of Hawking's black hole topology theorem to higher dimensions},
  Commun. Math. Phys. {\bf 266}, 571 (2006).

\bibitem{Garfinkle04}
  D. Garfinkle,
  {\it Gravitational collpase in anti de Sitter space},
  Phys. Rev. D {\bf 70}, 104015 (2004).

\bibitem{Goldwirth87}
  D.S. Goldwirth and T. Piran,
  {\it Gravitational collapse of massless scalar field and cosmic censorship}
  Phys. Rev. D {\bf 36}, 3575 (1987).

\bibitem{Gregory93}
  R. Gregory and R. Laflamme,
  {\it Black strings and p-branes are unstable},
  Phys. Rev. Lett. {\bf 70}, 2837 (1993).

\bibitem{Gundlach97}
  C. Gundlach,
  {\it Understanding critical collapse of a scalar field},
  Phys. Rev. D {\bf 55}, 695 (1997).

\bibitem{Gundlach07}
  C. Gundlach and J.M. Mart\'{\i}n-Garc\'{\i}a,
  {\it Critical phenomena in gravitational collapse},
  Living Rev. Rel. {\bf 10}, 5 (2007).

\bibitem{Gutperle04}
  M. Gutperle and P. Kraus,
  {\it Numerical study of cosmic censorship in string theory},
  J. High Energy Phys. {\bf 04}, 024 (2004).

\bibitem{Halliwell87}
  J.J. Halliwell,
  {\it Scalar fields in cosmology with an exponential potential},
  Phys. Lett. B {\bf 185}, 341 (1987).

\bibitem{Hawley00}
  S.H. Hawley and M.W. Choptuik,
  {\it Boson stars driven to the brink of black hole formation},
  Phys. Rev. D {\bf 62}, 104024 (2000).

\bibitem{Hertog04}
  T. Hertog, G.T. Horowitz, and K. Maeda,
  {\it Generic Cosmic-Censorship Violation in anti?de Sitter Space},
  Phys. Rev. Lett. {\bf 92}, 131101 (2004).

\bibitem{Hertog04a}
  T. Hertog, G.T. Horowitz, and K. Maeda,
  {\it Negative Energy in String Theory and Cosmic Censorship Violation},
  Phys. Rev. D {\bf 69}, 105001 (2004).

\bibitem{Hirschmann04}
  E.R. Hirschmann, A. Wang, and Y. Wu,
  {\it Collapse of a scalar field in 2+1 gravity},
  Class. Quantum Grav. {\bf 21}, 1791 (2004).

\bibitem{Honda02}
  E.P. Honda and M.W. Choptuik,
  {\it Fine structure of oscillons in the spherically symmetric $\varphi^4$ Klein-Gordon model}, Phys. Rev. D {\bf 65}, 084037 (2002).

\bibitem{Horowitz12}
  G.T. Horowitz ed.,
  {\it Black holes in higher dimensions},
  Cambridge University Press, Cambridge (2012).

\bibitem{Klainerman14}
  S. Klainerman, J. Luk, and I. Rodnianski,
  {\it A fully anisotropic mechanism for formation of trapped surfaces in vacuum},
  Invent. Math. {\bf 198}, 1 (2014).

\bibitem{Klainerman12}
  S. Klainerman and I. Rodnianski,
  {\it On the the formation of trapped surfaces},
  Acta Math. {\bf 208}, 211 (2012).

\bibitem{Langfelder05}
  P. Langfelder and R.B. Mann,
  {\it A note on spherically symmetric naked singularities in general dimension},
  Class. Quantum Grav. {\bf 22}, 1917 (2005).

\bibitem{Lehner10}
  L. Lehner and F. Pretorius,
  {\it Black Strings, Low Viscosity Fluids, and Violation of Cosmic Censorship},
  Phys. Rev. Lett.  {\bf 105}, 101102 (2010).

\bibitem{Li14}
  J. Li and P. Yu,
  {\it Construction of Cauchy data of vacuum Einstein field equations evolving to black holes},
  Ann. Math. {\bf 181}, 699 (2014).

\bibitem{Luk13}
  J. Luk and I. Rodnianski,
  {\it Nonlinear interactions of impulsive gravitational waves for the vacuum Einstein equations},
  arXiv:1301.1072 (2013).

\bibitem{Maartens10}
  R. Maartens and K. Koyama,
  {\it Brane-World Gravity},
  Living Rev. Rel. {\bf 13}, 5 (2010).

\bibitem{Martin03}
  J.M. Mart\'{\i}n-Garc\'{\i}a and C. Gundlach,
  {\it Global structure of Choptuik's critical solution in scalar field collapse},
  Phys. Rev. D {\bf 68}, 024011 (2003).

\bibitem{Mignemi89}
  S. Mignemi and D.L. Wiltshire,
  {\it Spherically symmetric solutions in dimensionally reduced spacetimes},
  Class. Quantum Grav. {\bf 6}, 987 (1989).

\bibitem{Overduin97}
  J.P. Overduin and P.S. Wesson,
  {\it Kaluza-Klein gravity},
  Phys. Rep. {\bf 283}, 303 (1997).

\bibitem{Paetz13}
  T. Paetz and W. Simon,
  {\it Marginally outer trapped surfaces in higher dimensions},
  Class. Quantum Grav.  {\bf 30}, 235005 (2013).

\bibitem{Penrose69}
  R. Penrose,
  {\it Gravitational collapse: The role of general relativity},
  Riv. Nuovo Cim. {\bf 1}, 252 (1969); Gen. Rel. Grav. {\bf 34}, 1141 (2002).

\bibitem{Penrose99}
  R. Penrose,
  {\it The Question of Cosmic Censorship},
  J. Astrophys. Astr. {\bf 20}, 233 (1999).

\bibitem{Poletti94}
  S.J. Poletti and D.L. Wiltshire,
  {\it Global properties of static spherically symmetric charged dilaton spacetimes with a Liouville potential},
  Phys. Rev. D {\bf 50}, 7260 (1994).

\bibitem{Reiterer11}
  M. Reiterer and E. Trubowitz,
  {\it Strongly focused gravitational waves},
  Comm. Math. Phys. {\bf 307}, 275 (2011).

\bibitem{Rendall05}
  A.D. Rendall,
  {\it The nature of spacetime singularities},
  arXiv:gr-qc/0503112 (2005).

\bibitem{Rendall11}
  A.D. Rendall and J.J.L. Vel\'{a}zquez,
  {\it A class of dust-like self-similar solutions of the massless Einstein-Vlasov system},
  Ann. Henri Poincar\'{e} {\bf 12}, 919 (2011).

\bibitem{Senovilla11}
  J.M.M. Senovilla,
  {\it Trapped surfaces},
  Int. J. Mod. Phys. D {\bf 20}, 2139 (2011).

\bibitem{Senovilla02}
  J.M.M. Senovilla,
  {\it Trapped surfaces, horizons and exact solutions in higher dimensions},
  Class. Quantum Grav. {\bf 19}, L113 (2002).

\bibitem{Townsend01}
  P.K. Townsend,
  {\it Quintessence from M-theory},
  J. High Energy Phys. {\bf 11}, 042 (2001).

\bibitem{Wald97}
  R.M. Wald,
  {\it Gravitational collapse and cosmic censorship},
  arXiv:gr-qc/9710068 (1997).

\bibitem{Wang03}
  A. Wang,
  {\it Critical collapse of a cylindrically symmetric scalar field in four-dimensional Einstein's theory of gravity},
  Phys. Rev. D {\bf 68}, 064006 (2003). 
  
\bibitem{Zhang15a}
  X. Zhang and H. L\"u,
  {\it Exact black hole formation in asymptotically (A)dS and flat spacetimes},
  Phys. Lett. B {\bf 736}, 455 (2014).

\bibitem{Zhang15b}
  X. Zhang and H. L\"u,
  {\it Critical behavior in a massless scalar field collapse with self-interacting potential},
  Phys. Rev. D {\bf 91} 044046 (2015).

\end{thebibliography}
\end{document}